\documentclass[a4paper,12pt]{article}

\usepackage{jheppub} 

\usepackage[T1]{fontenc} 

\title{\boldmath Bethe/Gauge Correspondence for \text{ABCDEFG}-type 3d Gauge Theories}

\author{Xiang-Mao Ding}
\author[1]{and Tinglyer Zhang \note{Corresponding author.}}

\affiliation{Institute of Applied Mathematics, Academy of Mathematics and Systems Science, \\
Chinese Academy of Sciences, Beijing 100190, China}

\emailAdd{xmding@amss.ac.cn}
\emailAdd{zhangting@amss.ac.cn}

\abstract{In this paper, we give a new effective superpotential that makes clear Bethe/Gauge correspondence between 2d (and 3d) $\text{SO/Sp}$ gauge theories and open $\text{XXX}$ (and $\text{XXZ}$) spin chains with diagonal boundary conditions, and also works in the case of 2d (and 3d) $\text{BC}_{N}$-type gauge theories which is not previously discussed in the literature. Especially, for exceptional Lie algebras $\text{F}_{4}$, $\text{G}_{2}$, we give the effective superpotential and vacuum equations. For $\text{E}_{6,7,8}$, we only give theirs effective superpotential for convenience.}


\begin{document} 
\maketitle
\flushbottom

\section{Introduction}
\label{sec:intro}

The dynamics of gauge theories is a long and fascinating subject, while the supersymmetric version is a newer one. Since the groundbreaking work of Seiberg and Witten \cite{SW94a,SW94b}, the integrable nature of supersymmetric gauge theories with eight super charges has gathered widespread interests. In recent years, it has been recognized that there exists an intimate connection between the vacua of the supersymmetric gauge theories and the quantum integrable systems \cite{NS09a,MNS00,GS08a,GS08b}. The effective twisted superpotential of a gauge theory corresponds to the Yang-Yang function of a quantum integrable system. In this regard, the space of supersymmetric vacua is equivalent to the state space of a quantum integrable system, whose Hamiltonian is the generator of the (twisted) chiral ring. That is to say, the spectrum of the quantum Hamiltonians coincides with the spectrum of the (twisted) chiral ring. Then the relationship was known as Bethe/Gauge correspondence, which opens a new door for the quantum integrable system and the gauge theories.

Quantum integrable systems and supersymmetric gauge field theories can be determined mutually. The low energy effective action of the gauge theory can be written if the corresponding classical integrable system is determined. Such as, the pure $\text{SU}(2)$ $\mathcal{N}=2$ super-Yang-Mills corresponds to the periodic $A_{N-1}$ Toda chain \cite{GKMMM}. At the same time, the effective twisted superpotential $W_{\text{eff}}(\sigma,m)$ (and the prepotential $\mathcal{F}(a)$) can be carried out using the gauge theory methods. For 4d gauge theory with the parameters $\varepsilon_{1}=\varepsilon$, $\varepsilon_{2}=0$ of the $\Omega$-bankground, the prepotential $\mathcal{F}(a)$ is played by the 2d twisted superpotential $W(a,\varepsilon)$, which is identified with the Yang-Yang function governing the spectrum of the quantum system. So the gauge theory can help to learn about the quantization of a classical integrable system \cite{NS09c}. What is more, there should be many applications of the correspondence \cite{NS09a,NS09b,MNS00,GS08a,GS08b}: the gauge theory applications, the study of quantum cohomology, (infinite dimensional) representation theory, harmonic analysis, the many-body quantum mechanics.

The Bethe/Gauge correspondence is a subject of research spanning over a decade \cite{NS09a,MNS00,GS08a,GS08b,NW10,NRS11} and even longer in the context of topological gauge theories \cite{Wit89,GN95,GN94a,GN94b}. This relationship became a subject of intense development after the seminal work where the instanton partition functions of the $\mathcal{S}$-class $\mathcal{N}=2$ gauge theories in the $\Omega$-background \cite{Nek04} were conjectured to be a Liouville (and, more generally, ADE type Toda theory) conformal blocks. In recent years, Bethe/Gauge correspondence used the techniques developed in the four dimensional instanton counting, which appeared in $\text{BPS/CFT}$ correspondence \cite{Nek16,Nek17a,Nek18,Nek19,Nek17b}, to calculate a quantum mechanical wave-function of a many-body system, or a spin chain \cite{NN21}. The Bethe/gauge correspondence can also be viewed as a correspondence between the supersymmetric gauge theories and representation theory of infinite dimensional algebras, especially a typical for the two-dimensional conformal field theories or integrable deformations thereof \cite{NVS18}.  

The correspondence between $\text{A}$-type gauge theories and closed $\text{XXX}$ spin chains has been carried out clearly in \cite{NS09a,NS09b}. The vacuum equations of 2d, 3d and 4d gauge theories correspond to rational, trigonometric and elliptic Bethe ansatz equations, respectively. For 2d, 3d $\text{BCD}$-type gauge groups, the correspondence has been proved partly in \cite{KZ21}. The correspondence worked perfectly for 2d gauge theories, but not as well in 3d case. To Excavate more the meaning of this duality, we will now give the explicit expression of the quantities we need to compare: the vacuum equation (in the Gauge side) and the Bethe ansatz equation (in the integrability side) for $\text{B}$ and $\text{D}$ type and clarify the problem for $\text{C}$ type. All the statements will also be clarified in the following sections. We can get the vacuum equation from the effective potential:
\begin{equation}\label{100}
    \text{exp}\left(\beta_{2}i \dfrac{\partial}{\partial \sigma}W^{3d}_{\text{eff}}(\sigma,m)\right)=1
\end{equation}
The following are the vacuum equation of 3d $\text{B}$-type gauge theory and the Bethe ansatz equation of open $\text{XXZ}$ spin chain with diagonal boundary condition \cite{KZ21}. The vacuum equation reads
\begin{equation}\label{4}
\begin{aligned}
&\dfrac{\text{sin}(\sigma_{j}-\beta_{2}\tilde{c})}{\text{sin}(\sigma_{j}+\beta_{2}\tilde{c})}\prod_{j\neq k}^{N}\dfrac{\text{sin}(\sigma_{j}\pm \sigma_{k}-\beta_{2}\tilde{c})}{\text{sin}(-\sigma_{j}\pm \sigma_{k}-\beta_{2}\tilde{c})}\prod_{a=1}^{N_{f}}\dfrac{\text{sin}(\sigma_{j}-m_{a}-\beta_{2}\tilde{c})}{\text{sin}(-\sigma_{j}-m_{a}-\beta_{2}\tilde{c})}=1
\end{aligned}
\end{equation}
where $\sigma_{i}$ are the eigenvalues of the complex scalar in the vector multiplet, $\beta_{2}$ is the $U(1)$ charge fugacity, $\tilde{c}$ is the rescaling $\text{R}$-charge of the scalar in the chiral multiplet and ${m}$ is the mass parameters, respectively. The Bethe ansatz equation is
\begin{equation}\label{23}
\begin{aligned}
&\dfrac{\text{sin}[\pi (u_{i}-\frac{\eta}{2}+\xi_{+})]}{\text{sin}[\pi(u_{i}+\frac{\eta}{2}-\xi_{+})]}\dfrac{\text{sin}[\pi (u_{i}-\frac{\eta}{2}+\xi_{-})]}{\text{sin}[\pi(u_{i}+\frac{\eta}{2}-\xi_{-})]}\\
&\times\prod_{a=1}^{L}\dfrac{\text{sin}[\pi (u_{i}+\frac{\eta}{2}+\eta s_{a}-\vartheta_{a})]\text{sin}[\pi(-u_{i}+\frac{\eta}{2}-\eta s_{a}-\vartheta_{a})]}{\text{sin}[\pi(-u_{i}+\frac{\eta}{2}+\eta s_{a}-\vartheta_{a})]\text{sin}[\pi(u_{i}+\frac{\eta}{2}-\eta s_{a}-\vartheta_{a})]}\\
&\times \prod_{j\neq i,j=1}^{M}\dfrac{\text{sin}[\pi (u_{j}+u_{i}-\eta)]\text{sin}[\pi(u_{j}-u_{i}-\eta)])}{\text{sin}[\pi(u_{j}-u_{i}+\eta)]\text{sin}[\pi(u_{j}+u_{i}+\eta)]}=1
\end{aligned}
\end{equation} 
where $u_{i}$ are spectral parameters, $\vartheta_{a}$ are inhomogeneous parameters, $\eta$ is the crossing parameter and $\xi_{\pm}$ are boundary parameters. $L$ is the length of the spin chain and $M$ is the number of excited states. The vacuum equation (\ref{4}) can be mapped to (\ref{23}) with the boundary condition chosen as
\begin{equation}\label{45}
\xi_{+}=\dfrac{\eta}{2},\quad \xi_{-}=-\dfrac{\eta}{2}
\end{equation}
We see that $N_{f}=2L$, $M \longleftrightarrow N$ and the mass parameters $\{m_{a}\}$ have to be paired as
\begin{equation}\label{7}
\begin{aligned}
&\pi u \longleftrightarrow \sigma ,\quad \pi \eta \longleftrightarrow \beta_{2}\tilde{c}\\
&\{-\pi \eta s_{a}-\dfrac{\pi \eta}{2} +\pi \vartheta_{a},-\pi \eta s_{a}+\dfrac{\pi \eta}{2} -\pi \vartheta_{a}\} \longleftrightarrow  m_{a}+\beta_{2}\tilde{c}
\end{aligned}
\end{equation} 
to make sure the duality. For $\text{D}$-type gauge theory, the vacuum equation is
\begin{equation}\label{6}
\prod_{j\neq k}^{N}\dfrac{\text{sin}(\sigma_{j}\pm \sigma_{k}-\beta_{2}\tilde{c})}{\text{sin}(-\sigma_{j}\pm \sigma_{k}-\beta_{2}\tilde{c})}\prod_{a=1}^{N_{f}}\dfrac{\text{sin}(\sigma_{j}-m_{a}-\beta_{2}\tilde{c})}{\text{sin}(-\sigma_{j}-m_{a}-\beta_{2}\tilde{c})}=1
\end{equation}
And it can be mapped to (\ref{23}) with boundary condition $\xi_{+}=\xi_{-}=i\infty$ with the dictionary (\ref{7}). The dictionary to map the vacuum equation to the Bethe equation is the same as the $\text{B}$-type above. For $\text{C}$-type gauge theory, the vacuum equation is
\begin{equation}\label{3}
\dfrac{\text{sin}^{2}(2\sigma_{j}-\beta_{2}\tilde{c})}{\text{sin}^{2}(2\sigma_{j}+\beta_{2}\tilde{c})}\prod_{j\neq k}^{N}\dfrac{\text{sin}(\sigma_{j}\pm \sigma_{k}-\beta_{2}\tilde{c})}{\text{sin}(-\sigma_{j}\pm \sigma_{k}-\beta_{2}\tilde{c})}\prod_{a=1}^{N_{f}}\dfrac{\text{sin}(\sigma_{j}-m_{a}-\beta_{2}\tilde{c})}{\text{sin}(-\sigma_{j}-m_{a}-\beta_{2}\tilde{c})}=1
\end{equation} 
We can see that $\xi$ has no suitable value to make the equations (\ref{23}) and (\ref{3}) dual. The barrier for type $\text{C}$ is the number $2$ of  $2\sigma$ in $\text{sin}^2(2\sigma_{i}\pm \beta_{2}\tilde{c})$. In this paper, we propose a solution to this problem: we change the representation of the gauge group and
find a new expression of the superpotential that can reproduce the results already found in \cite{NS09a, KZ21}. Furthermore, our expression is more general since it also works for the above C-type gauge theory, not discussed in the literature.

This paper is organized as follows. In section \ref{a}, we give a brief review on the integrability of the closed and open $\text{XXZ}$ spin chain. In section \ref{b}, we express the 3d $\mathcal{N}=2$ supersymmetric gauge theory on $D^{2}\times S^{1}$, and write down the new effective potential of different representations of Lie algebras. By using the new effective potential, we reproduce the well-known Bethe/Gauge correspondence between the vacuum equation of the $A$-type gauge theories and the Bethe ansatz equation of the closed $\text{XXZ}$ spin chains. In section \ref{c}, we give a new representation and define its corresponding superpotential. Then we extend the duality to the case of 3d (and 2d) $\text{BCD}$-type gauge theories and the open $\text{XXZ}$ (and $\text{XXX}$) spin chains with diagonal boundary condition, and 3d (and 2d) $\text{BC}_{N}$-type gauge theory and the open $\text{XXZ}$ (and $\text{XXX}$) spin chain with diagonal boundary condition. In section \ref{d}, we discuss the exceptional Lie algebras, in particular for $\text{F}_{4}$ and $\text{G}_{2}$ we gave the superpotential and the vacuum equations and for $\text{E}_{6,7,8}$ we gave the superpotential for convenience. Finally, we  give some meaning of our results and discuss some future directions in section \ref{e}.

\section{Spin chains and Bethe ansatz}\label{a}

In this section, we give a swift review of the integrable spin chains at the example of the $\text{XXZ}$ spin chain for $SU(2)$. The integrability of a spin chain is characterized by an $R$-matrix, $R(u):V\otimes V \rightarrow V\otimes V$, satisfying the Yang-Baxter equation,
\begin{equation}\label{1}
R_{12}(u-v)R_{13}(u)R_{23}(v)=R_{23}(v)R_{13}(u)R_{12}(u-v)
\end{equation}
where $u,v$ are called the spectral parameters. The object $R_{ij}$ are linear operators in the tensor product of the three linear space $V\otimes V\otimes V$ with $R_{12}=R(u)\otimes I$, $R_{23}=I\otimes R(u)$, etc. The most general $R$-matrix for a solvable $\text{XXZ}$ spin chain model can be expressed as \cite{Bax07}
\begin{equation}\label{2}
R^{\text{XXZ}}(u)=\begin{pmatrix}
[u+\eta]&0&0&0\\
0&[u]&[\eta]&0\\
0&[\eta]&[u]&0\\
0&0&0&[u+\eta]
\end{pmatrix}
\end{equation}
where 
\begin{equation}\label{}
[x]:=\dfrac{\text{sin}(\pi x)}{\text{sin}(\pi \eta)}
\end{equation}
where $\eta$ is the crossing parameter. It is easy to show that the above $R$-matrix satisfies the following relation \cite{WYC15}: $R(0)=P$, where $P$ is the permutation operator that acts as $P(x\otimes y)=y\otimes x$, $\forall x,y \in V$.

For a closed spin chain with periodic boundary condition, the monodromy matrix is
\begin{equation}\label{212}
T_{0}(u)=R_{0L}(u-\vartheta_{L})\cdots R_{01}(u-\vartheta_{1}):=\begin{pmatrix}
A(u)&B(u)\\
C(u)&D(u)
\end{pmatrix}
\end{equation}
where $T(u)\in \text{End}(V^{(0)}\otimes V^{\otimes L})$, $V^{(0)}$ is an auxiliary space, $A(u),B(u),C(u),D(u)\in \text{End}(V^{\otimes L})$, and {$\{\vartheta_{1},\cdots, \vartheta_{L}\}$ are called inhomogeneous parameters. The monodromy matrix satisfies the $\text{RTT}$-relation, 
\begin{equation}\label{203}
R_{12}(u-v)T_{1}(u)T_{2}(v)=T_{2}(v)T_{1}(u)R_{12}(u-v)
\end{equation}
which can be proven by substituting the definition of $T_{0}(u)$ in (\ref{212}) and using the Yang-Baxter equation (\ref{1}). The transfer matrix is given by
\begin{equation}
t(u)=\text{tr}_{0}T_{0}(u)=A(u)+D(u)
\end{equation}
From the $\text{RTT}$-relation (\ref{203}), the transfer matrices
corresponding to different spectral parameters commute \cite{Fad96}:
\begin{equation}
[t(u),t(v)]=0
\end{equation}
which ensures the integrability of this system. It is known that in the periodic spin chain case the Hamiltonian is given by the expression
\begin{equation}\label{205}
\mathcal{H}=\sum_{n=1}^{L-1}H_{n,n+1}
\end{equation}
where $H_{n,n+1}=P_{n,n+1}\frac{d}{du}R_{n,n+1}(u)|_{u=0}$ and $H_{L,L+1}=H_{L,1}$. The corresponding Hamiltonian for the closed $\text{XXZ}$-$\frac{1}{2}$ spin chain constructed from $(\ref{2})$ (with all inhomogeneous parameters turned off) is given by
\begin{equation}\label{}
    \mathcal{H}=\dfrac{1}{2}\sum_{n=1}^{L}\left(\sigma_{x}^{(n)}\sigma_{x}^{(n+1)}+\sigma_{y}^{(n)}\sigma_{y}^{(n+1)}+\text{cos}\pi \eta\sigma_{z}^{(n)}\sigma_{z}^{(n+1)}\right)
\end{equation}
where $\sigma_{x,y,z}^{(n)}$ are Pauli matrices assigned to the site $n$. We can assume the limit $\eta \rightarrow 0$ with the spectral parameter rescaled by $u\rightarrow u\eta$ to get the $\text{XXX}$ limit, where $[\eta]=\frac{\text{sin}(\pi \eta)}{\text{sin}(\pi \eta)}\rightarrow 1$, $[u]=\frac{\text{sin}(\pi u\eta)}{\text{sin}(\pi \eta)}\rightarrow u$ and $[u+\eta]=\frac{\text{sin}(\pi u\eta+\pi \eta)}{\text{sin}(\pi \eta)}\rightarrow u+1$. The twist is a transformation that preserves integrability (preserves the $\text{YBE}$ and the $\text{RTT}$ or $\text{RLL}$ relation). In order to preserve the integrability, we have freedom to multiply the monodromy matrix $T(u)$ by a matrix $U$ such that $[U\otimes U,R]=0$. For the case of $\text{XXX}$ spin chain, due to the symmetry, any arbitrary matrix $U$ works, but this is not true for $\text{XXZ}$. In the case of $\text{XXZ}$ spin chain, if we use the notation
\begin{equation}
    \iota(\theta)=\begin{pmatrix}
    1&0\\
    0&e^{i\theta}\\
    \end{pmatrix}
\end{equation}
then we can confirm that $\iota(\theta)\otimes \iota(\theta) $ commutes with the $R$-matrix,
\begin{equation}\label{204}
    [\iota(\theta)\otimes \iota(\theta),R^{\text{XXZ}}(u)]=0
\end{equation}
The transfer matrix with twisted periodic boundary condition $$\sigma_{x,y,z}^{L+1}=e^{\frac{i}{2}\theta \sigma_{z}}\sigma_{x,y,z}^{1}e^{-\frac{i}{2}\theta\sigma_{z}}$$ is given by
\begin{equation}\label{200}
t(u;\theta):=\text{tr}_{0}\iota_{0}(\theta)T_{0}(u)=A(u)+e^{i\theta}D(u)
\end{equation}
It can be checked that the state $\Omega=|\uparrow,\cdots,\uparrow \rangle$ with all spin up is the ground state of the system and it satisfies
\begin{equation}\label{}
A(u)\Omega=\delta_{+}(u)\Omega,\quad D(u)\Omega=\delta_{-}(u)\Omega,\quad C(u)\Omega=0
\end{equation}
where \begin{equation}\label{}
\delta_{+}(u)=\prod_{a=1}^{L}[u+\frac{\eta}{2}+\eta s_{a}-\vartheta_{a}],\qquad \delta_{-}(u)=\prod_{a=1}^{L}[u+\frac{\eta}{2}-\eta s_{a}-\vartheta_{a}]
\end{equation}
For a given ground state $\Omega$ of the system, the fact that
\begin{equation}\label{201}
\prod_{i=1}^{M}B(u_{i})\Omega
\end{equation}
is the Bethe ansatz state, which is motivated by the commutation relations of $[B,B]=0$ from the $\text{RTT}$ equation, with $M$ is the number of magnons in our spin chain and $u_{i}$ is the Bethe roots solution of the Bethe equation. Using the transfer matrix (\ref{200}) to act on the Bethe ansatz state (\ref{201}), and the commutation relation between $A(u)$, $D(u)$ and $B(u)$ from the $\text{RTT}$ relation, we get the Bethe ansatz equation with twisted periodic boundary condition
\begin{equation}\label{47}
\prod_{a=1}^{L}\dfrac{[u_{i}+\frac{\eta}{2}+\eta s_{a}-\vartheta_{a}]}{[u_{i}+\frac{\eta}{2}-\eta s_{a}-\vartheta_{a}]}=e^{i\theta}\prod_{j\neq i,j=1}^{M}\dfrac{[u_{i}-u_{j}+\eta]}{[u_{i}-u_{j}-\eta]}
\end{equation}
where we take spin $s_{a}$ representation at the $a$-th site.

Next, we review the derivation of  the Bethe ansatz equation of $\text{XXZ}$ spin chain with open boundary condition briefly. The Bethe ansatz equations have been derived in detail in \cite{KZ21}. In this paper, we focus on the open $\text{XXZ}$ spin chains with diagonal boundary operator  $K$-matrix \cite{IK94}
\begin{equation}\label{}
K^{\text{XXZ}}(u,\xi)=\begin{pmatrix}
[u+\xi]&0\\
0&-[u-\xi]
\end{pmatrix}
\end{equation}
The Hamiltonian of the open boundary can be found similarly as in (\ref{205}). The Hamiltonian of the open $\text{XXZ}$ spin chain with diagonal boundary conditions is \cite{Skl88}
\begin{equation}\label{}
\begin{aligned}
    \mathcal{H}=&\dfrac{1}{2}\sum_{i=1}^{L-1}\left(\sigma_{x}^{(i)}\sigma_{x}^{(i+1)}+\sigma_{y}^{(i)}\sigma_{y}^{(i+1)}+\text{cos}(\pi \eta)\sigma_{z}^{(i)}\sigma_{z}^{(i+1)}\right)\\
    &+\dfrac{\pi}{2}\text{cot}(\pi \xi_{-})\sigma_{z}^{(1)}+\dfrac{1}{2}\text{tan}(\pi \eta)\text{cot}(\pi \xi_{+})\sigma_{z}^{(L)}
\end{aligned}
\end{equation}
where $\xi_{-}$ and $\xi_{+}$ are the related left and right boundary parameter. The transfer matrix of an open chain is given \cite{WYC15}
\begin{equation}\label{}
t(u)=\text{Tr}_{0}K_{+}(u)T_{0}(u)K_{-}(u)T_{0}^{-1}(-u)
\end{equation} 
where {$T(u)$ is usually taken to be the same one in closed spin chain, $K_{-}(u)\in \text{End}(V^{(0)})$ and $K_{+}(u)\in \text{End}(V^{(0)})$ are the boundary operators}, which satisfy the Reflection Equation (RE) and dual RE, respectively,
\begin{equation}\label{}
\begin{aligned}
&R_{12}(u-v)K_{-}^{1}(u)R_{21}(u+v)K_{-}^{2}(v)\\
&=K_{-}^{2}(v)R_{21}(u+v)K_{-}^{1}(u)R_{12}(u-v)\\
\end{aligned}
\end{equation}
and
\begin{equation}
\begin{aligned}
&R_{12}(-u+v)K_{+}^{1}(u)R_{21}(-u-v-2\eta)K_{+}^{2}(v)\\
&=K_{+}^{2}(v)R_{21}(-u-v-2\eta)K_{+}^{1}(u)R_{12}(-u+v)
\end{aligned}
\end{equation}
where $K_{-}^{1}(u):=K_{-}(u)\otimes id_{V_{2}}$ and $K_{-}^{2}(u):=id_{V_{1}} \otimes K_{-}(u)$. Rather than the one-row monodromy matrix, we need to introduce the double-row monodromy matrix 
\begin{equation}\label{}
U_{-}(u)=T(u)K(u-\frac{\eta}{2},\xi_{-})\sigma_{y}T^{t}(-u)\sigma_{y}:=\begin{pmatrix}
\mathcal{A}(u)&\mathcal{B}(u)\\
\mathcal{C}(u)&\mathcal{D}(u)
\end{pmatrix}
\end{equation} 
The equivalent of the $\text{RTT}$ relation in case of open spin chain is the $\text{RURU}$ relation.
\begin{equation}\label{}
\begin{aligned}
&R_{12}(u-v)U_{-}^{1}(u)R_{12}(u+v-\eta)U_{-}^{2}(v)\\
&=U_{-}^{2}(v)R_{12}(u+v-\eta)U_{-}^{1}(u)R_{12}(u-v)
\end{aligned}
\end{equation}
Just like the closed spin chain, the ground state $\Omega$ is also the ground state of an open chain with diagonal boundary condition. The Bethe ansatz state is 
\begin{equation}\label{}
\prod_{i=1}^{M}\mathcal{B}(u_{i})\Omega
\end{equation}
Define $\tilde{\mathcal{D}}(u)=[2u]\mathcal{D}(u)-[\eta]\mathcal{A}(u)$. The transfer matrix $t(u)$ of the model with open boundary conditions is constructed by the double-row monodromy matrix as
\begin{equation}\label{}
\begin{aligned}
t(u)&=\text{Tr}\left(K_{+}(u+\eta,\xi_{+})U_{-}(u)\right)\\
&=\dfrac{[2u+\eta][u-\frac{\eta}{2}+\xi_{+}]}{[2u]}\mathcal{A}(u)-\dfrac{[u+\frac{\eta}{2}-\xi_{+}]}{[2u]}\tilde{\mathcal{D}}(u)
\end{aligned}
\end{equation}
where
\begin{equation}
 K_{+}(u,\xi_{+})=K(u+\frac{\eta}{2},\xi_{+}):=\begin{pmatrix}
 [u+\frac{\eta}{2}+\xi_{+}]&0\\
 0&-[u+\frac{\eta}{2}-\xi_{+}]
\end{pmatrix}
\end{equation}
For the excited Bethe states $\mathcal{B}(u_{1})\cdots \mathcal{B}(u_{M})\Omega$, we  consider
\begin{equation}\label{}
t(u)\prod_{i=1}^{M}\mathcal{B}(u_{i})\Omega=(X_{1}+Y_{1})\prod_{i=1}^{M}\mathcal{B}(u_{i})\Omega+(X_{2}+Y_{2})\mathcal{B}(u)\prod_{j\neq i}\mathcal{B}(u_{j})\Omega
\end{equation}
where
\begin{equation}\label{}
\begin{aligned}
X_{1}=&[2u+\eta][u-\frac{\eta}{2}+\xi_{+}][u-\dfrac{\eta}{2}+\xi_{-}]\prod_{i=1}^{M}\dfrac{[u-u_{i}-\eta][u+u_{i}-\eta]}{[u-u_{i}][u+u_{i}]}\delta_{+}(u)\delta_{-}(-u)\\
Y_{1}=&[2u-\eta][u+\frac{\eta}{2}-\xi_{-}][u+\frac{\eta}{2}-\xi_{+}]\prod_{i=1}^{M}\dfrac{[u-u_{i}+\eta][u+u_{i}+\eta]}{[u-u_{i}][u+u_{i}]}\delta_{+}(-u)\delta_{-}(u)\\
X_{2}=&\sum_{i=1}^{M}\dfrac{[2u+\eta][\eta][2u_{i}-\eta][2u][u_{i}-\frac{\eta}{2}+\xi_{+}][u_{i}-\frac{\eta}{2}+\xi_{-}]}{[]u_{i}-u][2u_{i}][u_{i}+u]}\\
&\times \prod_{j\neq i}\dfrac{[u_{i}+u_{j}-\eta][u_{i}-u_{j}-\eta]}{[u_{i}-u_{j}][u_{i}+u_{j}]}\delta_{+}(u_{i})\delta_{-}(-u_{i})\\
Y_{2}=&-\sum_{i=1}^{M}\dfrac{[2u+\eta][\eta][2u_{i}-\eta][2u][u_{i}+\frac{\eta}{2}-\xi_{+}][u_{i}+\frac{\eta}{2}-\xi_{-}]}{[u_{i}-u][2u_{i}][u_{i}+u]}\\
&\quad \times \prod_{j\neq i}\dfrac{[u_{i}+u_{j}-\eta][u_{i}-u_{j}-\eta]}{[u_{i}-u_{j}][u_{i}+u_{j}]}\delta_{+}(-u_{i})\delta_{-}(u_{i})
\end{aligned}
\end{equation}
If we want $X_{1}+Y_{1}$ to be the eigenvalue term and $\prod_{i=1}^{M}\mathcal{B}(u_{i})\Omega$ be the eigenstate, it is needed to put the unwanted term $X_{2}+Y_{2}=0$. From this last equality, we obtain the Bethe equations:
\begin{equation}\label{}
\begin{aligned}
&\dfrac{\text{sin}[\pi (u_{i}-\frac{\eta}{2}+\xi_{+})]}{\text{sin}[\pi(u_{i}+\frac{\eta}{2}-\xi_{+})]}\dfrac{\text{sin}[\pi (u_{i}-\frac{\eta}{2}+\xi_{-})]}{\text{sin}[\pi(u_{i}+\frac{\eta}{2}-\xi_{-})]}\\
&\times\prod_{a=1}^{L}\dfrac{\text{sin}[\pi (u_{i}+\frac{\eta}{2}+\eta s_{a}-\vartheta_{a})]\text{sin}[\pi(-u_{i}+\frac{\eta}{2}-\eta s_{a}-\vartheta_{a})]}{\text{sin}[\pi(-u_{i}+\frac{\eta}{2}+\eta s_{a}-\vartheta_{a})]\text{sin}[\pi(u_{i}+\frac{\eta}{2}-\eta s_{a}-\vartheta_{a})]}\\
&\times \prod_{j\neq i,j=1}^{M}\dfrac{\text{sin}[\pi (u_{j}+u_{i}-\eta)]\text{sin}[\pi(u_{j}-u_{i}-\eta])}{\text{sin}[\pi(u_{j}-u_{i}+\eta)]\text{sin}[\pi(u_{j}+u_{i}+\eta)]}=1
\end{aligned}
\end{equation}

\section{The gauge theory}\label{b}

In \cite{OR11}, Orlando and Reffert argue that the mathematical foundation of the Bethe/gauge correspondence is the geometric representation theory. Here we focus on a specific example of 2d $\mathcal{N}=(2,2)$ $U(N)$ gauge theory and $sl_{2}$ $\text{XXX}_\frac{1}{2}$ closed spin chains without inhomogeneity. On the gauge side, the low energy limit of the effective superpotential $W_{\text{eff}}^{2d}$ is the non-linear sigma model on $T^{*}Gr(N,L)$. Let $V$ be the fundamental representation of $sl_{2}$ and $V^{\otimes L}=\bigoplus_{N=0}^{L}V_{L-2N}$ be its $L$ times tensor product,  where $V_{L-2N}$ is $L-2N$ weight space.  The Hilbert space of the spin chain is $V^{\otimes L}$. Via the geometric representation theory, $V^{\otimes L}$ can be identified with $H_{*}[T^{*}(Gr(L))]$, which is the ground states of the non-linear sigma models on all the $T^{*}Gr(N,L)$ for $N=1,\cdots,L$. Thus, we identify the two space which stem from the two theories. The spectrum of the $N$ magnon sector is $V_{L-2N}$ is identified with $H_{*}[T^{*}(Gr(N,L))]$. Maybe geometric representation theory can helps us understand the underlying origins for Bethe/Gauge correspondence.

The disk partition function of 3d $\mathcal{N}= 2$ theory on $D^{2}\times S^{1}$ and 2d $\mathcal{N} = (2,2)$ theory on $D^{2}$ have been given in \cite{YS20}. The authors in \cite{KZ21} gave the effective twisted superpotential and vacuum equations of the gauge theories. In this section, different from the representation and the effective superpotential given in \cite{NS09a,KZ21}, we define a new effective superpotentials with giving a new representation. And we will describe these results in detail in subsection (\ref{f}).

\subsection{3d $\mathcal{N}$=2 theory}

For a gauge group $G$, we consider some linear representations $\mathcal{R}$ of $G$. For 2d $\mathcal{N}=(2,2)$ gauge theory, $G=U(N)$, if we considered the representation \cite{NS09a} 
\begin{equation}\label{99}
    \mathcal{R}=V\otimes V^{*}\otimes \mathcal{L}\oplus V\otimes \mathcal{F}\oplus V^{*}\otimes \tilde{\mathcal{F}}
\end{equation}
which corresponds to the $H^{\text{max}}=U(L)\times U(L)\times U(1)$ global symmetry group. Here $V=\mathbf{C}^{N}$ is the $N$-dimensional fundamental representation, $\mathcal{F}\approx \mathbf{C}^{L}$, $\tilde{\mathcal{F}}\approx \mathbf{C}^{L}$ are the $L$-dimensional fundamental representations of the first and second $U(L)$ factors in the flavour group, and $\mathcal{L}$ is the standard one-dimensional representation of the global group $U(1)$.  Let Euclidean space $\mathbf{E}$ to be the $(N-1)$-dimensional subspace of $R^{N}$ orthogonal to $\sigma_{1}+\cdots+\sigma_{N}$, the root system be the set of all vectors $\alpha\in \mathbf{E}$ for which $(\alpha,\alpha)=2$. It is obvious that the root system consists of all $\{\sigma_{i}-\sigma_{j}\}$, $i\neq j$ \cite{Hum72}. The corresponding effective twisted superpotential is
\begin{equation}\label{}
\begin{aligned}
W^{2d}_{\text{eff}}(\sigma)=&\sum_{j=1}^{N}\sum_{a=1}^{L}\left[(\sigma_{j}+m_{a}^{f})(\text{log}(\sigma_{j}+m_{a}^{f})-1)+(-\sigma_{i}+m_{a}^{\bar{f}})(\text{log}(-\sigma_{j}+m_{a}^{\bar{f}})-1)\right]\\
&+\sum_{j,k=1}^{N}(\sigma_{j}-\sigma_{k}+m_{\text{adj}})(\text{log}(\sigma_{j}-\sigma_{k}+m_{\text{adj}})-1)\\
&-2\pi i\sum_{j=1}^{N}(t+j-\frac{1}{2}(N+1))\sigma_{j}
\end{aligned}    
\end{equation}
where $\sigma_{j}$ are the eigenvalues of the complex scalar in the vector multiplet, $m$ is the twisted mass parameters, $t$ is a complex coupling number and the last term is the Fayet-Illiopoulos term, respectively. Actually, the vacuum equations for the two, three, and four dimensional $A$-type gauge theory can be summarized as:
\begin{equation}\label{88}
    \text{exp}\left(\dfrac{\partial W_{\text{eff}}
    (\sigma)}{\partial \sigma_{i}}\right)=1
\end{equation}
The vacuum equations of $\text{A}$-type 2d gauge theory are 
\begin{equation}\label{}
    \prod_{a=1}^{L}\dfrac{\sigma_{j}+m_{a}^{f}}{\sigma_{j}-m_{a}^{\bar{f}}}=-e^{2\pi it}\prod_{k=1}^{N}\dfrac{\sigma_{j}-\sigma_{k}-m_{\text{adj}}}{\sigma_{j}-\sigma_{k}+m_{\text{adj}}}, \qquad j=1,\cdots,N
\end{equation}

In \cite{KZ21}, the effective twisted superpotential with zero $\text{FI}$-term of 3d $\mathcal{N}$=2 theory on $D^{2}\times S^{1}$ has been calculated. We give a brief explanation about the computational progress. The partition function of 3d $\mathcal{N}$=2 theory on $D^{2}\times S^{1}$ has been carried out in \cite{YS20}, where the geometry of $D^{2}\times S^{1}$ is parameterized as
\begin{equation}\label{}
    ds^{2}=l^{2}(d\theta^{2}+r^{2}\text{sin}^{2}\theta d\phi^{2})+d\tau^{2},
\end{equation}
where the $S^{1}$ circle has a periodicity $\beta l$. The index on $D^{2}\times S^{1}$ is given by the following integral,
\begin{equation}\label{}
    \mathcal{I}=\dfrac{1}{|W_{G}|}\int \dfrac{d^{N}}{(2\pi)^{N}}e^{-S_{cl}}Z_{\text{vec}}Z_{\text{chi}}Z_{\text{bd}},
\end{equation}
where $W_{G}$ is the Weyl group of $G$. The key formula is the one-loop determinant of the vector multiplet
\begin{equation}\label{}
Z_{\text{vec}}=\prod_{\alpha\in \Delta}e^{\frac{1}{8\beta_{2}}(\alpha\cdot \sigma)^{2}}(e^{i\alpha\cdot \sigma};q^{2})_{\infty}
\end{equation}
with the set of the roots of $G$ denoted by $\Delta$, and the one-loop determinant of the chiral multiplet with Neumann boundary condition
\begin{equation}\label{}
    Z_{\text{chi}}^{\text{Neu}}=\prod_{w\in \mathcal{R}}e^{\mathcal{E}(iw\cdot \sigma+r\beta_{2}+im)}(e^{-iw\cdot \sigma-im}q^{r};q^{2})_{\infty}^{-1},\qquad q=e^{-\beta_{2}}
\end{equation}
here the weight's set of the corresponding representation denoted by $\mathcal{R}$, the $R$-charge of the scalar in the chiral multiplet $r$, and
\begin{equation}
    \mathcal{E}(x)=\dfrac{1}{8\beta_{2}}x^{2}-\dfrac{1}{4}x+\dfrac{\beta_{2}}{12}
\end{equation}
$\beta_{1}$ is the fugacity of the rotation along $S^{1}$, $\beta_{2}$ is the $U(1)$ charge fugacity, $\beta l=(\beta_{1}+\beta_{2})l$ is the circumference of $S^{1}$. The one-loop contribution of chiral multiplet with Dirichlet boundary condition is 
\begin{equation}\label{}
\begin{aligned}
Z_{\text{chi}}^{\text{Dir}}=\prod_{w\in \mathcal{R}}e^{\mathcal{E}(-iw\cdot \sigma+(2-r)\beta_{2}-im)}(e^{iw\cdot \sigma+im}q^{2-r};q^{2})_{\infty},\qquad q=e^{-\beta_{2}}
\end{aligned}
\end{equation}
The relation between the partition function and the effective potential is 
\begin{equation}\label{}
    \mathcal{I}\sim \text{exp}\left(\frac{1}{\epsilon}W_{\text{eff}}(\sigma,m)\right)
\end{equation}
In fact, we only need the two one-loop determinants $Z_{\text{vec}}$ and $Z_{\text{chi}}$ in $\mathcal{I}$ to get $W_{\text{eff}}^{3d}$. With the rescaling $\beta_{2}\rightarrow \beta_{2}\epsilon/2$ and $r\rightarrow 1+\tfrac{2i}{\epsilon}\tilde{c}$, the effective potential is obtained by taking limit $\epsilon \rightarrow 0$
\begin{equation}\label{98}
\begin{aligned}
W^{3d}_{\text{eff}}(\sigma,m)=&\dfrac{1}{\beta_{2}}\sum_{w \in \mathcal{R}}\sum_{a=1}^{N_{f}}\text{Li}_{2}(e^{-iw\cdot \sigma-im_{a}-i\beta_{2}\tilde{c}})-\dfrac{1}{4\beta_{2}}\sum_{w\in \mathcal{R}}\sum_{a=1}^{N_{f}}(w\cdot \sigma+m_{a}+\beta_{2}\tilde{c})^{2}\\
&-\dfrac{1}{\beta_{2}}\sum_{\alpha \in \Delta}\text{Li}_{2}(e^{i\alpha\cdot \sigma})+\dfrac{1}{4\beta_{2}}\sum_{\alpha\in \Delta}(\alpha\cdot \sigma)^{2}
\end{aligned}
\end{equation}
where all the chiral multiplets are put to satisfy the Neumann boundary condition. The $\text{Li}_{2}(z)$ is called the dilogarithm,
\begin{equation}
\text{Li}_{2}(z)=\sum_{k=1}^{\infty}\dfrac{z^{k}}{k^{2}}
\end{equation}
 The correspondence between 2d (or 3d) gauge theories with $\text{SO/Sp}$ gauge groups and $\text{XXX}$ (or $\text{XXZ}$) spin chains with open boundary conditions
are studied in \cite{KZ21}, and the Bethe/Gauge correspondence have been 
proved for $\text{SO}$-type gauge theory. For $\text{C}$-type gauge theory, only the Bethe/Gauge correspondence between 2d $\mathcal{N}=(2,2)$ gauge theory and $\text{XXX}$ open spin chain is obtained. The corresponding vacuum equations of different gauge groups are dual to different open boundary conditions of the $\text{XXZ}$ spin chain. And they also modified the vacuum equation (\ref{88}) as in \cite{KZ21}
\begin{equation}\label{}
    \text{exp}\left(\beta_{2}i \dfrac{\partial}{\partial \sigma}W^{3d}_{\text{eff}}(\sigma,m)\right)=1
\end{equation}
An important identity is given by
\begin{equation}\label{}
    \text{exp}\left(\dfrac{\partial}{\partial x}\text{Li}_{2}(e^{\pm x})\right)=(1-e^{\pm x})^{\mp}
\end{equation}

\subsection{Refined representation: the new superpotential}\label{f}
 
Nekrasov and Shatashvili use matter field which is in chiral multiplets and gauge field which is in vector multiplets in \cite{NS09a} to compute effective twisted superpotential $W_{\text{eff}}(\sigma,m)$. We see that in the above 3d $\mathcal{N}=2$ gauge theory,  the effective potential just includes one vector multiplet and one chiral multiplet. Simply, we can regard that the above representations just include one adjoint representation and one fundamental or anti-fundamental representation. In fact, the representation $\mathcal{R}$ of $\text{BCD}$-type gauge group in \cite{KZ21} can be written as
\begin{equation}\label{}
    \mathcal{R}=V\otimes V^{*}\oplus V\otimes \mathcal{F}
\end{equation}
where $V$ is the fundamental representation and $\mathcal{F}$ is the $N_{f}$ dimensional representation.  If  we change the representation $\mathcal{R}$ to a new representation $\mathcal{R}^{'}$ liking (\ref{99}):
\begin{equation}\label{}
    \mathcal{R}^{'}=V\otimes V^{*}\oplus V\otimes \mathcal{F}\oplus V\otimes \mathcal{F}^{'}
\end{equation}
the corresponding effective potential of 3d $\mathcal{N}=2$ gauge theory with a zero $\text{FI}$-term is given by 
\begin{equation}\label{}
    \begin{aligned}
    W^{3d}_{\text{eff}}(\sigma,m)= &\dfrac{1}{\beta_{2}}\sum_{w \in \mathcal{R}}\sum_{a=1}^{N_{f}}\text{Li}_{2}(e^{-iw\cdot \sigma-im_{a}-i\beta_{2}\tilde{c}})-\dfrac{1}{4\beta_{2}}\sum_{w\in \mathcal{R}}\sum_{a=1}^{N_{f}}(w\cdot \sigma+m_{a}+\beta_{2}\tilde{c})^{2}\\
    &\dfrac{1}{\beta_{2}}\sum_{w \in \mathcal{R}}\sum_{a=1}^{N_{f}^{'}}\text{Li}_{2}(e^{iw\cdot \sigma-im_{a}-i\beta_{2}\tilde{c}})-\dfrac{1}{4\beta_{2}}\sum_{w\in \mathcal{R}}\sum_{a=1}^{N_{f}^{'}}(w\cdot \sigma-m_{a}-\beta_{2}\tilde{c})^{2}\\
    &-\dfrac{1}{\beta_{2}}\sum_{\alpha \in \Delta}\text{Li}_{2}(e^{i\alpha\cdot \sigma})+\dfrac{1}{4\beta_{2}}\sum_{\alpha\in \Delta}(\alpha\cdot \sigma)^{2}
\end{aligned}
\end{equation}
And the new vacuum equation is the same to (\ref{100}). In this paper, for a specific supersymmetry gauge theory and known Lie group, we change the above effective superpotential (\ref{98}) to
\begin{equation}\label{97}
    \begin{aligned}
    W^{3d}_{\text{eff}}(\sigma,m)=&\dfrac{1}{\beta_{2}}\sum_{w \in \mathcal{R}}\sum_{a=1}^{N_{f}}\frac{4}{\alpha_{i}^{2}}\text{Li}_{2}(e^{-iw\cdot \sigma-im_{a}-i\beta_{2}\tilde{c}})-\dfrac{1}{4\beta_{2}}\sum_{w\in \mathcal{R}}\sum_{a=1}^{N_{f}}\frac{4}{\alpha_{i}^{2}}(w\cdot \sigma+m_{a}+\beta_{2}\tilde{c})^{2}\\
    &-\dfrac{1}{\beta_{2}}\sum_{\alpha \in \Delta}\frac{4}{\alpha_{i}^{2}}\text{Li}_{2}(e^{i\alpha\cdot \sigma})+\dfrac{1}{4\beta_{2}}\sum_{\alpha\in \Delta}\frac{4}{\alpha_{i}^{2}}(\alpha\cdot \sigma)^{2}
   \end{aligned}
\end{equation}
where $\alpha_{i}$ is the length of the roots of the Lie algebra of the Lie groups. We can rewrite the formula (\ref{97}) as
\begin{equation}\label{48}
    \begin{aligned}
    W^{3d}_{\text{eff}}(\sigma,m)=&\dfrac{1}{\beta_{2}}\sum_{w \in \mathcal{R}}\sum_{a=1}^{N_{f}}\text{Li}_{2}(e^{-iw\cdot \sigma-im_{a}-i\beta_{2}\tilde{c}})-\dfrac{1}{4\beta_{2}}\sum_{w\in \mathcal{R}}\sum_{a=1}^{N_{f}}(w\cdot \sigma+m_{a}+\beta_{2}\tilde{c})^{2}\\
    &\dfrac{1}{\beta_{2}}\sum_{w \in \mathcal{R}}\sum_{a=1}^{N_{f}^{'}}\text{Li}_{2}(e^{iw\cdot \sigma-im_{a}^{'}-i\beta_{2}\tilde{c}})-\dfrac{1}{4\beta_{2}}\sum_{w\in \mathcal{R}}\sum_{a=1}^{N_{f}^{'}}(w\cdot \sigma-m_{a}^{'}+\beta_{2}\tilde{c})^{2}\\
      &-\dfrac{1}{\beta_{2}}\sum_{\alpha \in \Delta}\frac{4}{\alpha_{i}^{2}}\text{Li}_{2}(e^{i\alpha\cdot \sigma})+\dfrac{1}{4\beta_{2}}\sum_{\alpha\in \Delta}\frac{4}{\alpha_{i}^{2}}(\alpha\cdot \sigma)^{2}
   \end{aligned}
\end{equation}
The two formulas (\ref{97}) and (\ref{48}) are the same in terms of calculation. They change the one adjoint chiral multiplet to $\frac{4}{\alpha_{i}^{2}}$ times adjoint chiral multiplets and keep $N_{f}=N_{f}^{'}$. When we consider different gauge group, the effective superpotential is invariable. The only difference is the root system of different Lie algebra. The factor $\frac{4}{\alpha_{i}^{2}}$ in effective superpotential $W^{3d}_{\text{eff}}$ (\ref{97}) expands the range of application of (\ref{98}).

As an example, we calculate $\text{A}$-type gauge theory. In 3d $\mathcal{N}=2$ $U(N)$ gauge theory, we choose a new representation

\begin{equation}
    \mathcal{R}=V\otimes V^{*}\oplus V\otimes \mathcal{F}\oplus V\otimes \mathcal{F}\oplus V^{*}\otimes \tilde{\mathcal{F}}\oplus V^{*}\otimes \tilde{\mathcal{F}}
\end{equation}
which corresponds to the theory with the $H^{\text{max}}=U(L)\times U(L)\times U(L)\times U(L)$ global symmetry group. Similarly, $V=\mathbf{C}^{N}$ is the $N$-dimensional fundamental representation, $\mathcal{F}\approx \mathbf{C}^{L}$, $\tilde{\mathcal{F}}\approx \mathbf{C}^{L}$ are the $L$-dimensional fundamental representations. The effective superpotential for this theory is given by

\begin{equation}\label{}
\begin{aligned}
W^{3d}_{\text{eff}}=&\dfrac{2}{\beta_{2}}\sum_{j\neq k}^{N}\text{Li}_{2}(e^{-i(\sigma_{j}-\sigma_{k})-im_{\text{adj}}})-\dfrac{1}{2\beta_{2}}\sum_{j\neq k}^{N}(\sigma_{j}-\sigma_{k}+m_{\text{adj}})^{2}\\
&+\dfrac{2}{\beta_{2}}\sum_{j=1}^{N}\sum_{a=1}^{N_{f}}\text{Li}_{2}(e^{-i \sigma_{j}-im_{a}})-\dfrac{1}{2\beta_{2}}\sum_{j=1}^{N}\sum_{a=1}^{N_{f}}( \sigma_{j}+m_{a})^{2}\\
&+\dfrac{2}{\beta_{2}}\sum_{j=1}^{N}\sum_{a=1}^{N_{f}^{'}}\text{Li}_{2}(e^{i \sigma_{j}-im_{a}^{'}})-\dfrac{1}{2\beta_{2}}\sum_{j=1}^{N}\sum_{a=1}^{N_{f}^{'}}( \sigma_{j}-m_{a}^{'})^{2}\\
&-\dfrac{2}{\beta_{2}}\sum_{j\neq k}^{N}\text{Li}_{2}(e^{i(\sigma_{j}-\sigma_{k})})+\dfrac{1}{2\beta_{2}}\sum_{j\neq k}^{N}(\sigma_{j}-\sigma_{k})^{2}
\end{aligned}
\end{equation}
where we have absorbed $\beta_{2}\tilde{c}$ into the mass parameters. Then the vacuum equation is written by

\begin{equation}\label{49}
\dfrac{\prod_{a=1}^{N_{f}}\text{sin}^{2}(\sigma_{j}-m_{a}^{'})}{\prod_{a=1}^{N_{f}^{'}}\text{sin}^{2}(\sigma_{j}+m_{a})}=\prod_{k\neq j}^{N}\dfrac{\text{sin}^{2}(\sigma_{j}-\sigma_{k}+m_{\text{adj}})}{\text{sin}^{2}(\sigma_{j}-\sigma_{k}-m_{\text{adj}})}
\end{equation}
According to the assumption $N_{f}=N_{f}^{'}$, we can take the square root of the vacuum equation (\ref{49}) to get two type vacuum equations equivalently

\begin{equation}\label{46}
\dfrac{\prod_{a=1}^{N_{f}}\text{sin}(\sigma_{j}-m_{a}^{'})}{\prod_{a=1}^{N_{f}^{'}}\text{sin}(\sigma_{j}+m_{a})}=\prod_{k\neq j}^{N}\dfrac{\text{sin}(\sigma_{j}-\sigma_{k}+m_{\text{adj}})}{\text{sin}(\sigma_{j}-\sigma_{k}-m_{\text{adj}})}
\end{equation}
and

\begin{equation}\label{202}
\dfrac{\prod_{a=1}^{N_{f}}\text{sin}(\sigma_{j}-m_{a}^{'})}{\prod_{a=1}^{N_{f}^{'}}\text{sin}(\sigma_{j}+m_{a})}=-\prod_{k\neq j}^{N}\dfrac{\text{sin}(\sigma_{j}-\sigma_{k}+m_{\text{adj}})}{\text{sin}(\sigma_{j}-\sigma_{k}-m_{\text{adj}})}
\end{equation}
We can regard the vacuum equations (\ref{46}) and (\ref{202}) as two independent connecting branches in the moduli space of vacuum of the correspondence gauge theory. In our construction, a vacuum equation of gauge theory will be dual to a pair of Bethe vacuum equations, they appear in the meantime. In our case, the Bethe/Gauge correspondence is just choosing one of them. The choice is the same for all Lie algebra cases for our consideration in this paper, we will not mention this point 
repeatedly in the following. The dictionary is given by
\begin{equation}\label{}
\begin{aligned}
&\pi u \longleftrightarrow \sigma ,\quad \pi \eta \longleftrightarrow m_{\text{adj}}\\
&M \longleftrightarrow N,\quad L\longleftrightarrow N_{f}=N_{f}^{'}\\
&\{-\pi \eta s_{a}-\dfrac{\pi \eta}{2} +\pi \vartheta_{a}\}\longleftrightarrow m_{a}^{'},\quad \{-\pi \eta s_{a}+\dfrac{\pi \eta}{2} -\pi \vartheta_{a}\} \longleftrightarrow  m_{a}
\end{aligned}
\end{equation}
to match the Bethe ansatz equation (\ref{47}) with vacuum equation (\ref{46}). So the vacuum equation of $\text{A}$-type theory in \cite{NS09a} is a special case of our vacuum equation exactly. The correspondence here certainly works in parallel after taking the 2d limit $\beta_{2}\sim \beta_{2}\rightarrow 0$ in the gauge theory and the $\text{XXX}$ limit, $[u]\rightarrow u$, of the spin chain.
\section{Bethe/Gauge correspondence}\label{c} 
Let us recall the general effective potential expression (\ref{48})
\begin{equation}\label{}
    \begin{aligned}
    W^{3d}_{\text{eff}}(\sigma,m)=  &-\dfrac{1}{\beta_{2}}\sum_{\alpha \in \Delta}\frac{4}{\alpha_{i}^{2}}\text{Li}_{2}(e^{i\alpha\cdot \sigma})+\dfrac{1}{4\beta_{2}}\sum_{\alpha\in \Delta}\frac{4}{\alpha_{i}^{2}}(\alpha\cdot \sigma)^{2}\\
        &+\dfrac{1}{\beta_{2}}\sum_{w \in \mathcal{R}}\sum_{a=1}^{N_{f}}\text{Li}_{2}(e^{-iw\cdot \sigma-im_{a}-i\beta_{2}\tilde{c}})-\dfrac{1}{4\beta_{2}}\sum_{w\in \mathcal{R}}\sum_{a=1}^{N_{f}}(w\cdot \sigma+m_{a}+\beta_{2}\tilde{c})^{2}\\
    &+\dfrac{1}{\beta_{2}}\sum_{w \in \mathcal{R}}\sum_{a=1}^{N_{f}^{'}}\text{Li}_{2}(e^{iw\cdot \sigma-im_{a}^{'}-i\beta_{2}\tilde{c}})-\dfrac{1}{4\beta_{2}}\sum_{w\in \mathcal{R}}\sum_{a=1}^{N_{f}^{'}}(w\cdot \sigma-m_{a}^{'}-\beta_{2}\tilde{c})^{2}
   \end{aligned}
\end{equation}
We use the same process to calculate the effective potential of $\text{BCD}$-type gauge theories. Then we discuss the connection between the gauge theory results and the open spin chains models based on this expression. 

\subsection{$\text{B}_{N}$-type gauge theory}

In the case of $SO(2N+1)$ gauge group, all the roots are given by $\{\pm \sigma_{i}\pm \sigma_{j}\}$ for all the possible combinations of $i<j$ and $\{\pm \sigma_{i}\}_{i=1}^{N}$.The effective superpotential is 
\begin{equation*}
	\begin{aligned}
	W^{3d}_{\text{eff}}(\sigma,m)=&-\dfrac{2}{\beta_{2}}\sum_{j<k}^{N}\text{Li}_{2}(e^{i(\pm\sigma_{j}\pm \sigma_{k})})+\dfrac{1}{2\beta_{2}}\sum_{j<k}^{N}(\pm \sigma_{j}\pm \sigma_{k})^{2}\\
	&+\dfrac{1}{\beta_{2}}\sum_{j<k}^{N}\text{Li}_{2}(e^{-i(\pm \sigma_{j}\pm \sigma_{k})-im_{\text{adj}}})-\dfrac{1}{4\beta_{2}}\sum_{j<k}^{N}(\pm \sigma_{j}\pm \sigma_{k}+m_{\text{adj}})^{2}\\
	&+\dfrac{1}{\beta_{2}}\sum_{j=1}^{N}\sum_{a=1}^{N_{f}}\text{Li}_{2}(e^{-(\pm i \sigma_{j}+im_{a})})-\dfrac{1}{4\beta_{2}}\sum_{j=1}^{N}\sum_{a=1}^{N_{f}}(\pm \sigma_{j}+m_{a})^{2}\\
	\end{aligned}
\end{equation*}
\begin{equation}\label{}
\begin{aligned}
\qquad \qquad \qquad \quad&-\dfrac{4}{\beta_{2}}\sum_{j=1}^{N}\text{Li}_{2}(e^{\pm i \sigma_{j}})+\dfrac{1}{\beta_{2}}\sum_{j=1}^{N}\sigma_{j}^{2}\\
&+\dfrac{1}{\beta_{2}}\sum_{j<k}^{N}\text{Li}_{2}(e^{-i(\pm \sigma_{j}\pm \sigma_{k})-im_{\text{adj}}})-\dfrac{1}{4\beta_{2}}\sum_{j<k}^{N}(\pm \sigma_{j}\pm \sigma_{k}+m_{\text{adj}})^{2}\\
&+\dfrac{4}{\beta_{2}}\sum_{j=1}^{N}\text{Li}_{2}(e^{(\pm i \sigma_{j}-im_{\text{adj}})})-\dfrac{1}{\beta_{2}}\sum_{j=1}^{N}(\sigma_{j}\pm m_{\text{adj}})^{2}\\
&+\dfrac{1}{\beta_{2}}\sum_{j=1}^{N}\sum_{a=1}^{N_{f}^{'}}\text{Li}_{2}(e^{\pm i \sigma_{j}-im_{a}^{'}})-\dfrac{1}{4\beta_{2}}\sum_{j=1}^{N}\sum_{a=1}^{N_{f}^{'}}(\pm \sigma_{j}-m_{a}^{'})^{2}
\end{aligned}
\end{equation}
and the vacuum equation is given by
\begin{equation}\label{17}
\begin{aligned}
&\dfrac{\text{sin}^{4}(\sigma_{j}-m_{\text{adj}})}{\text{sin}^{4}(\sigma_{j}+m_{\text{adj}})}\prod_{j\neq k}^{N}\dfrac{\text{sin}(\sigma_{j}\pm \sigma_{k}-m_{\text{adj}})}{\text{sin}(-\sigma_{j}\pm \sigma_{k}-m_{\text{adj}})}\prod_{a=1}^{N_{f}}\dfrac{\text{sin}(\sigma_{j}-m_{a})}{\text{sin}(-\sigma_{j}-m_{a})}\\
&\times \prod_{j\neq k}^{N}\dfrac{\text{sin}(\sigma_{j}\pm \sigma_{k}-m_{\text{adj}})}{\text{sin}(-\sigma_{j}\pm \sigma_{k}-m_{\text{adj}})}\prod_{a=1}^{N_{f}^{'}}\dfrac{\text{sin}(\sigma_{j}-m_{a}^{'})}{\text{sin}(-\sigma_{j}-m_{a}^{'})}=1
\end{aligned}
\end{equation}
In the case $N_{f}=N_{f}^{'}$, we can rewrite the vacuum equation (\ref{17}) as 
\begin{equation}\label{}
\left[\dfrac{\text{sin}^{2}(\sigma_{j}-m_{\text{adj}})}{\text{sin}^{2}(\sigma_{j}+m_{\text{adj}})}\prod_{j\neq k}^{N}\dfrac{\text{sin}(\sigma_{j}\pm \sigma_{k}-m_{\text{adj}})}{\text{sin}(-\sigma_{j}\pm \sigma_{k}-m_{\text{adj}})}\prod_{a=1}^{N_{f}}\dfrac{\text{sin}(\sigma_{j}-m_{a})}{\text{sin}(-\sigma_{j}-m_{a})}\right]^{2}=1
\end{equation}
Then we get two type vacuum equations by taking the square root of (\ref{17})
\begin{equation}\label{5}
\dfrac{\text{sin}^{2}(\sigma_{j}-m_{\text{adj}})}{\text{sin}^{2}(\sigma_{j}+m_{\text{adj}})}\prod_{j\neq k}^{N}\dfrac{\text{sin}(\sigma_{j}\pm \sigma_{k}-m_{\text{adj}})}{\text{sin}(-\sigma_{j}\pm \sigma_{k}-m_{\text{adj}})}\prod_{a=1}^{N_{f}}\dfrac{\text{sin}(\sigma_{j}-m_{a})}{\text{sin}(-\sigma_{j}-m_{a})}=1
\end{equation}
and
\begin{equation}\label{206}
\dfrac{\text{sin}^{2}(\sigma_{j}-m_{\text{adj}})}{\text{sin}^{2}(\sigma_{j}+m_{\text{adj}})}\prod_{j\neq k}^{N}\dfrac{\text{sin}(\sigma_{j}\pm \sigma_{k}-m_{\text{adj}})}{\text{sin}(-\sigma_{j}\pm \sigma_{k}-m_{\text{adj}})}\prod_{a=1}^{N_{f}}\dfrac{\text{sin}(\sigma_{j}-m_{a})}{\text{sin}(-\sigma_{j}-m_{a})}=-1
\end{equation}
Compared to the general Bethe ansatz equations (\ref{23}) for the open spin chain with diagonal boundary conditions, the vacuum equation (\ref{5}) can be mapped to the above Bethe ansatz equation (\ref{23}) with the boundary conditions
\begin{equation}\label{}
\xi_{+}=\xi_{-}=-\dfrac{\eta}{2}
\end{equation}
We also see the fact that $N_{f}=N_{f}^{'}=2L$ is an even integer and the map is
\begin{equation}\label{9}
\begin{aligned}
&\pi u \longleftrightarrow \sigma ,\quad \pi \eta \longleftrightarrow m_{\text{adj}}\\
& M \longleftrightarrow N,\quad 2L\longleftrightarrow N_{f} \\
&\{-\pi \eta s_{a}-\dfrac{\pi \eta}{2} +\pi \vartheta_{a},-\pi \eta s_{a}+\dfrac{\pi \eta}{2} -\pi \vartheta_{a}\} \longleftrightarrow  m_{a}
 \end{aligned}
\end{equation}
to sure the duality. If we choose $s_{1}=s_{2}=\dfrac{1}{2}$ and $\vartheta_{1}=\vartheta_{2}=2k$, $k\in \mathbb{Z}$, then 
\begin{equation}
\dfrac{\text{sin}[\pi (u_{i}+\frac{\eta}{2}+\eta s_{1}-\vartheta_{1})]}{\text{sin}[\pi(-u_{i}+\frac{\eta}{2}+\eta s_{1}-\vartheta_{1})]}=\dfrac{\text{sin}[\pi(u_{i}+\pi\eta)]}{\text{sin}[\pi(-u_{i}+\pi\eta)]},\quad \dfrac{\text{sin}[\pi(-u_{i}+\frac{\eta}{2}-\eta s_{2}-\vartheta_{2})]}{\text{sin}[\pi(u_{i}+\frac{\eta}{2}-\eta s_{2}-\vartheta_{2})]}=-1
\end{equation}
The corresponding $m_{1}\longleftrightarrow -\pi \eta+ 2k\pi$, $m_{2}\longleftrightarrow -2k\pi$ and
 \begin{equation}
 \dfrac{\text{sin}(\sigma_{j}-m_{1})}{\text{sin}(-\sigma_{j}-m_{1})}=\dfrac{\text{sin}(\sigma_{j}+\pi \eta)}{\text{sin}(-\sigma_{j}+\pi\eta)},\quad  \dfrac{\text{sin}(\sigma_{j}-m_{2})}{\text{sin}(-\sigma_{j}-m_{2})}=-1
\end{equation}
In this way, the factor $\frac{\text{sin}^{2}(\sigma_{j}-m_{\text{adj}})}{\text{sin}^{2}(\sigma_{j}+m_{\text{adj}})}$ in (\ref{5}) can be reduced as
 \begin{equation}
 \dfrac{\text{sin}^{2}(\sigma_{j}-m_{\text{adj}})}{\text{sin}^{2}(\sigma_{j}+m_{\text{adj}})}\dfrac{\text{sin}(\sigma_{j}-m_{1})}{\text{sin}(-\sigma_{j}-m_{1})}\dfrac{\text{sin}(\sigma_{j}-m_{2})}{\text{sin}(-\sigma_{j}-m_{2})}=\dfrac{\text{sin}(\sigma_{j}-m_{\text{adj}})}{\text{sin}(\sigma_{j}+m_{\text{adj}})}
\end{equation}
Then the vacuum equation (\ref{5}) is the same as the vacuum equation (\ref{4}) in \cite{KZ21}. And the dictionary is 
\begin{equation}\label{}
\begin{aligned}
&\pi u \longleftrightarrow \sigma ,\quad \pi \eta \longleftrightarrow m_{\text{adj}}\\
& M\longleftrightarrow N,\quad 2L\longleftrightarrow N_{f}-2\\
&\{-\pi \eta s_{a}-\dfrac{\pi \eta}{2} +\pi \vartheta_{a},-\pi \eta s_{a}+\dfrac{\pi \eta}{2} -\pi \vartheta_{a}\} \longleftrightarrow  m_{a}
 \end{aligned}
\end{equation} 
We can see the vacuum equation (\ref{4}) is a special case of the vacuum equation (\ref{5}).

\subsection{$\text{C}_{N}$-type gauge theory}

For $\text{Sp}(N)$ gauge theory, all the roots are given by $\{\pm \sigma_{i}\pm \sigma_{j}\}$ for $i<j$ and $\{\pm 2\sigma_{i}\}_{i=1}^{N}$. The effective superpotential is
\begin{equation*}
	\begin{aligned}
	W^{3d}_{\text{eff}}(\sigma,m)=&-\dfrac{2}{\beta_{2}}\sum_{j<k}^{N}\text{Li}_{2}(e^{i(\pm\sigma_{j}\pm \sigma_{k})})+\dfrac{1}{2\beta_{2}}\sum_{j<k}^{N}(\pm \sigma_{j}\pm \sigma_{k})^{2}\\
&+\dfrac{1}{\beta_{2}}\sum_{j<k}^{N}\text{Li}_{2}(e^{-i(\pm \sigma_{j}\pm \sigma_{k})-im_{\text{adj}}})-\dfrac{1}{4\beta_{2}}\sum_{j<k}^{N}(\pm \sigma_{j}\pm \sigma_{k}+m_{\text{adj}})^{2}\\
&+\dfrac{1}{\beta_{2}}\sum_{j=1}^{N}\sum_{a=1}^{N_{f}}\text{Li}_{2}(e^{-(\pm i \sigma_{j}+im_{a})})-\dfrac{1}{4\beta_{2}}\sum_{j=1}^{N}\sum_{a=1}^{N_{f}}(\pm \sigma_{j}+m_{a})^{2}\\
&-\dfrac{1}{\beta_{2}}\sum_{j=1}^{N}\text{Li}_{2}(e^{\pm 2i \sigma_{j}})+\dfrac{1}{\beta_{2}}\sum_{j=1}^{N}\sigma_{j}^{2}\\
	\end{aligned}
\end{equation*}	 
\begin{equation}\label{}
\begin{aligned}
\qquad \qquad \qquad \ &+\dfrac{1}{\beta_{2}}\sum_{j=1}^{N}\text{Li}_{2}(e^{\pm 2i \sigma_{j}-im_{\text{adj}}})-\dfrac{1}{4\beta_{2}}\sum_{j=1}^{N}(2\sigma_{j}\pm m_{\text{adj}})^{2}\\
&+\dfrac{1}{\beta_{2}}\sum_{j<k}^{N}\text{Li}_{2}(e^{-i(\pm \sigma_{j}\pm \sigma_{k})-im_{\text{adj}}})-\dfrac{1}{4\beta_{2}}\sum_{j<k}^{N}(\pm \sigma_{j}\pm \sigma_{k}+m_{\text{adj}})^{2}\\
&+\dfrac{1}{\beta_{2}}\sum_{j=1}^{N}\sum_{a=1}^{N_{f}^{'}}\text{Li}_{2}(e^{\pm i \sigma_{j}-im_{a}^{'}})-\dfrac{1}{4\beta_{2}}\sum_{j=1}^{N}\sum_{a=1}^{N_{f}^{'}}(\pm \sigma_{j}-m_{a}^{'})^{2}
\end{aligned}
\end{equation}
and the vacuum equation is given by
\begin{equation}\label{18}
\begin{aligned}
&\dfrac{\text{sin}^{2}(2\sigma_{j}-m_{\text{adj}})}{\text{sin}^{2}(2\sigma_{j}+m_{\text{adj}})}\prod_{j\neq k}^{N}\dfrac{\text{sin}(\sigma_{j}\pm \sigma_{k}-m_{\text{adj}})}{\text{sin}(-\sigma_{j}\pm \sigma_{k}-m_{\text{adj}})}\prod_{a=1}^{N_{f}}\dfrac{\text{sin}(\sigma_{j}-m_{a})}{\text{sin}(-\sigma_{j}-m_{a})}\\
&\times \prod_{j\neq k}^{N}\dfrac{\text{sin}(\sigma_{j}\pm \sigma_{k}-m_{\text{adj}})}{\text{sin}(-\sigma_{j}\pm \sigma_{k}-m_{\text{adj}})}\prod_{a=1}^{N_{f}^{'}}\dfrac{\text{sin}(\sigma_{j}-m_{a}^{'})}{\text{sin}(-\sigma_{j}-m_{a}^{'})}=1
\end{aligned}
\end{equation}
Using the following formula
\begin{equation}\label{}
\dfrac{\text{sin}^{2}(2\sigma_{i}-m_{\text{adj}})}{\text{sin}^{2}(2\sigma_{i}+m_{\text{adj}})}=\dfrac{\text{sin}^{2}(\sigma_{i}-\frac{m_{\text{adj}}}{2})\text{cos}^{2}(\sigma_{i}-\frac{m_{\text{adj}}}{2})}{\text{sin}^{2}(\sigma_{i}+\frac{m_{\text{adj}}}{2})\text{cos}^{2}(\sigma_{i}+\frac{m_{\text{adj}}}{2})}
\end{equation}
With $N_{f}=N_{f}^{'}$, the vacuum equation can be written as two type vacuum equations after taking the square root of (\ref{18})
\begin{equation}\label{207}
\dfrac{\text{sin}(\sigma_{j}-\frac{m_{\text{adj}}}{2})\text{cos}(\sigma_{j}-\frac{m_{\text{adj}}}{2})}{\text{sin}(\sigma_{j}+\frac{m_{\text{adj}}}{2})\text{cos}(\sigma_{j}+\frac{m_{\text{adj}}}{2})}\prod_{j\neq k}^{N}\dfrac{\text{sin}(\sigma_{j}\pm \sigma_{k}-m_{\text{adj}})}{\text{sin}(-\sigma_{j}\pm \sigma_{k}-m_{\text{adj}})}\prod_{a=1}^{N_{f}}\dfrac{\text{sin}(\sigma_{j}-m_{a})}{\text{sin}(-\sigma_{j}-m_{a})}=1
\end{equation}
and
\begin{equation}\label{8}
\dfrac{\text{sin}(\sigma_{j}-\frac{m_{\text{adj}}}{2})\text{cos}(\sigma_{j}-\frac{m_{\text{adj}}}{2})}{\text{sin}(\sigma_{j}+\frac{m_{\text{adj}}}{2})\text{cos}(\sigma_{j}+\frac{m_{\text{adj}}}{2})}\prod_{j\neq k}^{N}\dfrac{\text{sin}(\sigma_{j}\pm \sigma_{k}-m_{\text{adj}})}{\text{sin}(-\sigma_{j}\pm \sigma_{k}-m_{\text{adj}})}\prod_{a=1}^{N_{f}}\dfrac{\text{sin}(\sigma_{j}-m_{a})}{\text{sin}(-\sigma_{j}-m_{a})}=-1
\end{equation}
The same dictionary (\ref{9}) maps the vacuum equation (\ref{8}) to the Bethe ansatz equation (\ref{23}) with the following boundary conditions:
\begin{itemize}
\item \qquad $\xi_{+}=\dfrac{1}{2},\qquad \xi_{-}=0$
\item \qquad $\xi_{+}=0,\qquad \xi_{-}=\dfrac{1}{2}$
\end{itemize}
Due to the periodicity of the trigonometric function $\text{sin}$ and $\text{cos}$, there are infinite kinds of boundary parameters in the Bethe ansatz equations (\ref{23}) to match the vacuum equations of the $\text{C}_{N}$-type gauge theory. We can get other boundary condition by using $\xi$ to add k with k integer. And we can choose one of them to get the Bethe/Gauge correspondence. If we regard the two boundary operator $K_{\pm}$ as the same, there are one type of boundary condition, actually. Up to now, we can say that we solved the problem of the factor $\text{sin}^{2}(2\sigma_{j}-\beta_{2}\tilde{c})$ in \cite{KZ21}.

\subsection{$\text{D}_{N}$-type gauge theory}

In the case of $\text{SO}(2N)$ gauge group, all the roots are given by $\{\pm \sigma_{i}\pm \sigma_{j}\}$ for all $1\le i< j\le N$. The effective superpotential is  
 \begin{equation}\label{}
\begin{aligned}
W^{3d}_{\text{eff}}(\sigma,m)=&-\dfrac{2}{\beta_{2}}\sum_{j<k}^{N}\text{Li}_{2}(e^{i(\pm\sigma_{j}\pm \sigma_{k})})+\dfrac{1}{2\beta_{2}}\sum_{j<k}^{N}(\pm \sigma_{j}\pm \sigma_{k})^{2}\\
&+\dfrac{1}{\beta_{2}}\sum_{j<k}^{N}\text{Li}_{2}(e^{-i(\pm \sigma_{j}\pm \sigma_{k})-im_{\text{adj}}})-\dfrac{1}{4\beta_{2}}\sum_{j<k}^{N}(\pm \sigma_{j}\pm \sigma_{k}+m_{\text{adj}})^{2}\\
&+\dfrac{1}{\beta_{2}}\sum_{j=1}^{N}\sum_{a=1}^{N_{f}}\text{Li}_{2}(e^{-(\pm i \sigma_{j}+im_{a})})-\dfrac{1}{4\beta_{2}}\sum_{j=1}^{N}\sum_{a=1}^{N_{f}}(\pm \sigma_{j}+m_{a})^{2}\\
&+\dfrac{1}{\beta_{2}}\sum_{j<k}^{N}\text{Li}_{2}(e^{-i(\pm \sigma_{j}\pm \sigma_{k})-im_{\text{adj}}})-\dfrac{1}{4\beta_{2}}\sum_{j<k}^{N}(\pm \sigma_{j}\pm \sigma_{k}+m_{\text{adj}})^{2}\\
&+\dfrac{1}{\beta_{2}}\sum_{j=1}^{N}\sum_{a=1}^{N_{f}^{'}}\text{Li}_{2}(e^{\pm i \sigma_{j}-im_{a}^{'}})-\dfrac{1}{4\beta_{2}}\sum_{j=1}^{N}\sum_{a=1}^{N_{f}^{'}}(\pm \sigma_{j}-m_{a}^{'})^{2}
\end{aligned}
\end{equation}
and the vacuum equation is given by
\begin{equation}\label{19}
\begin{aligned}
&\prod_{j\neq k}^{N}\dfrac{\text{sin}(\sigma_{j}\pm \sigma_{k}-m_{\text{adj}})}{\text{sin}(-\sigma_{j}\pm \sigma_{k}-m_{\text{adj}})}\prod_{a=1}^{N_{f}}\dfrac{\text{sin}(\sigma_{j}-m_{a})}{\text{sin}(-\sigma_{j}-m_{a})}\\
&\times \prod_{j\neq k}^{N}\dfrac{\text{sin}(\sigma_{j}\pm \sigma_{k}-m_{\text{adj}})}{\text{sin}(-\sigma_{j}\pm \sigma_{k}-m_{\text{adj}})}\prod_{a=1}^{N_{f}^{'}}\dfrac{\text{sin}(\sigma_{j}-m_{a}^{'})}{\text{sin}(-\sigma_{j}-m_{a}^{'})}=1
\end{aligned}
\end{equation}
Using $N_{f}=N_{f}^{'}$, we also can rewrite the vacuum equation as two type vacuum equations after taking the square root of (\ref{19})
\begin{equation}\label{10}
\prod_{j\neq k}^{N}\dfrac{\text{sin}(\sigma_{j}\pm \sigma_{k}-m_{\text{adj}})}{\text{sin}(-\sigma_{j}\pm \sigma_{k}-m_{\text{adj}})}\prod_{a=1}^{N_{f}}\dfrac{\text{sin}(\sigma_{j}-m_{a})}{\text{sin}(-\sigma_{j}-m_{a})}=1
\end{equation}
and
\begin{equation}\label{208}
\prod_{j\neq k}^{N}\dfrac{\text{sin}(\sigma_{j}\pm \sigma_{k}-m_{\text{adj}})}{\text{sin}(-\sigma_{j}\pm \sigma_{k}-m_{\text{adj}})}\prod_{a=1}^{N_{f}}\dfrac{\text{sin}(\sigma_{j}-m_{a})}{\text{sin}(-\sigma_{j}-m_{a})}=-1
\end{equation}
With the same map (\ref{9}), we can see that the vacuum equation (\ref{10}) can be mapped to the Bethe ansatz equation (\ref{23}) with the boundary condition chosen as
\begin{equation}
\xi_{+}=\xi_{-}=i\infty
\end{equation}
For $\text{D}$-type gauge theory,  the correspondence is exactly the same as the results in \cite{KZ21}. So the vacuum equation (\ref{6}) is a special case of our vacuum equation.

\subsection{$\text{BC}_{N}$-type gauge theory}
Now we consider a new 3d $\text{BC}_{N}$ gauge theory \cite{Mac03}. All the roots are given by $\{\pm \sigma_{i}\pm \sigma_{j}\}$ for all the possible combinations of $1\le i<j\le N$, $\{\pm \sigma_{i}\}_{i=1}^{N}$ and $\{\pm 2\sigma_{i}\}_{i=1}^{N}$. The effective superpotential is 
\begin{equation}\label{}
\begin{aligned}
W^{3d}_{\text{eff}}(\sigma,m)=&-\dfrac{2}{\beta_{2}}\sum_{j<k}^{N}\text{Li}_{2}(e^{i(\pm\sigma_{j}\pm \sigma_{k})})+\dfrac{1}{2\beta_{2}}\sum_{j<k}^{N}(\pm \sigma_{j}\pm \sigma_{k})^{2}\\
&+\dfrac{2}{\beta_{2}}\sum_{j<k}^{N}\text{Li}_{2}(e^{-i(\pm \sigma_{j}\pm \sigma_{k})-im_{\text{adj}}})-\dfrac{1}{4\beta_{2}}\sum_{j<k}^{N}(\pm \sigma_{j}\pm \sigma_{k}+m_{\text{adj}})^{2}\\
&+\dfrac{1}{\beta_{2}}\sum_{j=1}^{N}\sum_{a=1}^{N_{f}}\text{Li}_{2}(e^{-(\pm i \sigma_{j}+im_{a})})-\dfrac{1}{4\beta_{2}}\sum_{j=1}^{N}\sum_{a=1}^{N_{f}}(\pm \sigma_{j}+m_{a})^{2}\\
&-\dfrac{4}{\beta_{2}}\sum_{j=1}^{N}\text{Li}_{2}(e^{\pm i \sigma_{j}})+\dfrac{1}{\beta_{2}}\sum_{j=1}^{N}\sigma_{j}^{2}\\
&+\dfrac{4}{\beta_{2}}\sum_{j=1}^{N}\text{Li}_{2}(e^{\pm i \sigma_{j}-im_{\text{adj}}})-\dfrac{1}{\beta_{2}}\sum_{j=1}^{N}(\sigma_{j}\pm m_{\text{adj}})^{2}\\
&-\dfrac{1}{\beta_{2}}\sum_{i=1}^{N}\text{Li}_{2}(e^{\pm 2 i\sigma_{j}})+\dfrac{1}{\beta_{2}}\sum_{j=1}^{N}\sigma_{j}^{2}\\
&+\dfrac{1}{\beta_{2}}\sum_{j=1}^{N}\text{Li}_{2}(e^{\pm 2i \sigma_{j}-im_{\text{adj}}})-\dfrac{1}{\beta_{2}}\sum_{j=1}^{N}(2\sigma_{j}\pm m_{\text{adj}})^{2}\\
&+\dfrac{1}{\beta_{2}}\sum_{j=1}^{N}\sum_{a=1}^{N_{f}^{'}}\text{Li}_{2}(e^{\pm i \sigma_{j}-im_{a}^{'}})-\dfrac{1}{4\beta_{2}}\sum_{j=1}^{N}\sum_{a=1}^{N_{f}^{'}}(\pm \sigma_{j}-m_{a}^{'})^{2}
\end{aligned}
\end{equation}
and the vacuum equation is given by
\begin{equation}\label{11}
\begin{aligned}
&\dfrac{\text{sin}^{4}(\sigma_{j}-m_{\text{adj}})}{\text{sin}^{4}(\sigma_{j}+m_{\text{adj}})}\prod_{j\neq k}^{N}\dfrac{\text{sin}(\sigma_{j}\pm \sigma_{k}-m_{\text{adj}})}{\text{sin}(-\sigma_{j}\pm \sigma_{k}-m_{\text{adj}})}\prod_{a=1}^{N_{f}}\dfrac{\text{sin}(\sigma_{j}-m_{a})}{\text{sin}(-\sigma_{j}-m_{a})}\\
&\times\dfrac{\text{sin}^{2}(2\sigma_{j}-m_{\text{adj}})}{\text{sin}^{2}(2\sigma_{j}+m_{\text{adj}})}\prod_{j\neq k}^{N}\dfrac{\text{sin}(\sigma_{j}\pm \sigma_{k}-m_{\text{adj}})}{\text{sin}(-\sigma_{j}\pm \sigma_{k}-m_{\text{adj}})}\prod_{a=1}^{N_{f}^{'}}\dfrac{\text{sin}(\sigma_{j}-m_{a}^{'})}{\text{sin}(-\sigma_{j}-m_{a}^{'})}=1
\end{aligned}
\end{equation}
Noticing that $N_{f}=N_{f}^{'}$, we also can rewrite the vacuum equation as two type vacuum equations after taking the square root (\ref{11})
\begin{equation}\label{209}
\begin{aligned}
&\dfrac{\text{sin}(\sigma_{j}-\frac{m_{\text{adj}}}{2})\text{cos}(\sigma_{j}-\frac{m_{\text{adj}}}{2})}{\text{sin}(\sigma_{j}+\frac{m_{\text{adj}}}{2})\text{cos}(\sigma_{j}+\frac{m_{\text{adj}}}{2})}\prod_{j\neq k}^{N}\dfrac{\text{sin}(\sigma_{j}\pm \sigma_{k}-m_{\text{adj}})}{\text{sin}(-\sigma_{j}\pm \sigma_{k}-m_{\text{adj}})}\\
&\times \dfrac{\text{sin}^{2}(\sigma_{j}-m_{\text{adj}})}{\text{sin}^{2}(\sigma_{j}+m_{\text{adj}})}\prod_{a=1}^{N_{f}}\dfrac{\text{sin}(\sigma_{j}-m_{a})}{\text{sin}(-\sigma_{j}-m_{a})}=1
\end{aligned}
\end{equation}
and
\begin{equation}\label{12}
\begin{aligned}
&\dfrac{\text{sin}(\sigma_{j}-\frac{m_{\text{adj}}}{2})\text{cos}(\sigma_{j}-\frac{m_{\text{adj}}}{2})}{\text{sin}(\sigma_{j}+\frac{m_{\text{adj}}}{2})\text{cos}(\sigma_{j}+\frac{m_{\text{adj}}}{2})}\prod_{j\neq k}^{N}\dfrac{\text{sin}(\sigma_{j}\pm \sigma_{k}-m_{\text{adj}})}{\text{sin}(-\sigma_{j}\pm \sigma_{k}-m_{\text{adj}})}\\
&\times \dfrac{\text{sin}^{2}(\sigma_{j}-m_{\text{adj}})}{\text{sin}^{2}(\sigma_{j}+m_{\text{adj}})}\prod_{a=1}^{N_{f}}\dfrac{\text{sin}(\sigma_{j}-m_{a})}{\text{sin}(-\sigma_{j}-m_{a})}=-1
\end{aligned}
\end{equation}
If we choose $s_{1}=s_{2}=s_{3}=s_{4}=\dfrac{1}{2}$ and $\vartheta_{1}=\vartheta_{2}=\vartheta_{3}=\vartheta_{4}=2k$, $k\in \mathbb{Z}$, then we get the corresponding $m_{1}=m_{3} \longleftrightarrow -\pi \eta + 2k\pi$, $m_{2}=m_{4} \longleftrightarrow -2k\pi$ and
 \begin{equation}
 \dfrac{\text{sin}(\sigma_{j}-m_{1})}{\text{sin}(-\sigma_{j}-m_{1})}=\dfrac{\text{sin}(\sigma_{j}+\pi \eta)}{\text{sin}(-\sigma_{j}+\pi\eta)},\quad  \dfrac{\text{sin}(\sigma_{j}-m_{2})}{\text{sin}(-\sigma_{j}-m_{2})}=-1
\end{equation}
\begin{equation}
 \dfrac{\text{sin}(\sigma_{j}-m_{3})}{\text{sin}(-\sigma_{j}-m_{3})}=\dfrac{\text{sin}(\sigma_{j}+\pi \eta)}{\text{sin}(-\sigma_{j}+\pi\eta)},\quad  \dfrac{\text{sin}(\sigma_{j}-m_{4})}{\text{sin}(-\sigma_{j}-m_{4})}=-1
\end{equation}
The factor $\frac{\text{sin}(\sigma_{j}-\frac{m_{\text{adj}}}{2})\text{cos}(\sigma_{j}-\frac{m_{\text{adj}}}{2})}{\text{sin}(\sigma_{j}+\frac{m_{\text{adj}}}{2})\text{cos}(\sigma_{j}+\frac{m_{\text{adj}}}{2})}\frac{\text{sin}^{2}(\sigma_{j}-m_{\text{adj}})}{\text{sin}^{2}(\sigma_{j}+m_{\text{adj}})}$ of vacuum equation (\ref{12}) can be changed as
\begin{equation}
\begin{aligned}
&\dfrac{\text{sin}(\sigma_{j}-\frac{m_{\text{adj}}}{2})\text{cos}(\sigma_{j}-\frac{m_{\text{adj}}}{2})}{\text{sin}(\sigma_{j}+\frac{m_{\text{adj}}}{2})\text{cos}(\sigma_{j}+\frac{m_{\text{adj}}}{2})}\dfrac{\text{sin}^{2}(\sigma_{j}-m_{\text{adj}})}{\text{sin}^{2}(\sigma_{j}+m_{\text{adj}})}\\
&\times \dfrac{\text{sin}(\sigma_{j}-m_{1})}{\text{sin}(-\sigma_{j}-m_{1})}\dfrac{\text{sin}(\sigma_{j}-m_{2})}{\text{sin}(-\sigma_{j}-m_{2})} \dfrac{\text{sin}(\sigma_{j}-m_{3})}{\text{sin}(-\sigma_{j}-m_{3})}\dfrac{\text{sin}(\sigma_{j}-m_{4})}{\text{sin}(-\sigma_{j}-m_{4})}\\
&=\dfrac{\text{sin}(\sigma_{j}-\frac{m_{\text{adj}}}{2})\text{cos}(\sigma_{j}-\frac{m_{\text{adj}}}{2})}{\text{sin}(\sigma_{j}+\frac{m_{\text{adj}}}{2})\text{cos}(\sigma_{j}+\frac{m_{\text{adj}}}{2})}
\end{aligned}
\end{equation}
With this equation, the vacuum equation (\ref{12}) can be mapped to the Bethe ansatz equation (\ref{23}) with the boundary condition
\begin{itemize}
\item \qquad $\xi_{+}=\dfrac{1}{2},\qquad \xi_{-}=0$
\item \qquad $\xi_{+}=0,\qquad \xi_{-}=\dfrac{1}{2}$
\end{itemize}
And the dictionary is changed into 
\begin{equation}\label{}
\begin{aligned}
&\pi u \longleftrightarrow \sigma ,\quad \pi \eta \longleftrightarrow m_{\text{adj}}\\
&M\longleftrightarrow N,\quad 2L\longleftrightarrow N_{f}-4\\
&\{-\pi \eta s_{a}-\dfrac{\pi \eta}{2} +\pi \vartheta_{a},-\pi \eta s_{a}+\dfrac{\pi \eta}{2} -\pi \vartheta_{a}\} \longleftrightarrow  m_{a}
 \end{aligned}
\end{equation} 
Like the $\text{C}$-type gauge theory, there are infinite type of boundary condition on account of the trigonometric function $\text{sin}$ and $\text{cos}$. The sites of fixed boundary parameters in open spin chain are arbitrary. As we know that, the Dynkin diagram of $\text{B}$-type and $\text{BC}_{N}$ can be obtained from $\text{A}$-type with appropriate folding. From this point of view, the folding structure of $\text{B}$-type has two nodes, while the folding structure of $\text{BC}_{N}$ has four nodes. There are irregular behaviours on the nodes. We see that, the nodes numbers are just coincide with the numbers of fixed boundary parameters. The coincidence is worthwhile to clarify further from the representation theory.

\subsection{2d limits}

In the $2d$ limit, the $\text{XXZ}$ spin chain is changed to $\text{XXX}$ spin chain, where $[u]\rightarrow u$, it is very straightforward to get Bethe/Gauge correspondence. Firstly, the Bethe ansatz equations (\ref{23}) degenerate to
\begin{equation}\label{13}
\begin{aligned}
&\dfrac{(u_{j}-\frac{\eta}{2}+\xi_{+})}{(u_{j}+\frac{\eta}{2}-\xi_{+})}\dfrac{(u_{j}-\frac{\eta}{2}+\xi_{-})}{(u_{j}+\frac{\eta}{2}-\xi_{-})}\prod_{a=1}^{L}\dfrac{(u_{j}+\frac{\eta}{2}+\eta s_{a}-\vartheta_{a})(-u_{j}+\frac{\eta}{2}-\eta s_{a}-\vartheta_{a})}{(-u_{j}+\frac{\eta}{2}+\eta s_{a}-\vartheta_{a})(u_{j}+\frac{\eta}{2}-\eta s_{a}-\vartheta_{a})}\\
&\times \prod_{k\neq j,k=1}^{M}\dfrac{ (u_{k}-u_{j}-\eta)(u_{k}+u_{j}-\eta)}{(u_{k}-u_{j}+\eta)(u_{k}+u_{j}+\eta)}=1
\end{aligned}
\end{equation} 
The vacuum equation of 3d supersymmetry gauge theory also degenerate to vacuum equation of 2d supersymmetry gauge theory. For 2d $\text{D}$-type theory, the vacuum equation is given by
\begin{equation}\label{20}
\begin{aligned}
\prod_{j\neq k}^{N}\dfrac{(\sigma_{j}\pm \sigma_{k}-m_{\text{adj}})}{(-\sigma_{j}\pm \sigma_{k}-m_{\text{adj}})}\prod_{a=1}^{N_{f}}\dfrac{(\sigma_{j}-m_{a})}{(-\sigma_{j}-m_{a})}\prod_{j\neq k}^{N}\dfrac{(\sigma_{j}\pm \sigma_{k}-m_{\text{adj}})}{(-\sigma_{j}\pm \sigma_{k}-m_{\text{adj}})}\prod_{a=1}^{N_{f}^{'}}\dfrac{(\sigma_{j}-m_{a}^{'})}{(-\sigma_{j}-m_{a}^{'})}=1
\end{aligned}
\end{equation}
Taking the square root of vacuum equation (\ref{20}), we get two type vacuum equation with $N_{f}=N_{f}^{'}$
\begin{equation}\label{14}
\prod_{j\neq k}^{N}\dfrac{(\sigma_{j}\pm \sigma_{k}-m_{\text{adj}})}{(-\sigma_{j}\pm \sigma_{k}-m_{\text{adj}})}\prod_{a=1}^{N_{f}}\dfrac{(\sigma_{j}-m_{a})}{(-\sigma_{j}-m_{a})}=1
\end{equation}
and 
\begin{equation}\label{}
\prod_{j\neq k}^{N}\dfrac{(\sigma_{j}\pm \sigma_{k}-m_{\text{adj}})}{(-\sigma_{j}\pm \sigma_{k}-m_{\text{adj}})}\prod_{a=1}^{N_{f}}\dfrac{(\sigma_{j}-m_{a})}{(-\sigma_{j}-m_{a})}=-1
\end{equation}
Comparing the Bethe ansatz equation (\ref{13}) with the vacuum equation (\ref{14}), we choose the boundary conditions
\begin{equation}
\xi_{+}=\xi_{-}=i\infty
\end{equation}
to get Bethe/Gauge correspondence. The precise relation reads as follows:
\begin{equation}\label{15}
\begin{aligned}
&u \longleftrightarrow \sigma ,\quad \eta \longleftrightarrow m_{\text{adj}}\\
& M \longleftrightarrow N,\quad 2L\longleftrightarrow N_{f}\\
&\{-\eta s_{a}-\dfrac{\eta}{2} +\vartheta_{a},-\eta s_{a}+\dfrac{ \eta}{2} -\vartheta_{a}\} \longleftrightarrow  m_{a}
 \end{aligned}
\end{equation}
In particular, the vacuum equation of 2d $\text{D}$-type supersymmetry gauge theory in \cite{KZ21} is the same as (\ref{14}). So we can see their result is a special case of ours. 

For $\text{C}$-type theory,  the vacuum equation is 
\begin{equation}\label{21}
\begin{aligned}
&\dfrac{(2\sigma_{j}-m_{\text{adj}})^{2}}{(2\sigma_{j}+m_{\text{adj}})^{2}}\prod_{j\neq k}^{N}\dfrac{(\sigma_{j}\pm \sigma_{k}-m_{\text{adj}})}{(-\sigma_{j}\pm \sigma_{k}-m_{\text{adj}})}\prod_{a=1}^{N_{f}}\dfrac{(\sigma_{j}-m_{a})}{(-\sigma_{j}-m_{a})}\\
&\times\prod_{j\neq k}^{N}\dfrac{(\sigma_{j}\pm \sigma_{k}-m_{\text{adj}})}{(-\sigma_{j}\pm \sigma_{k}-m_{\text{adj}})}\prod_{a=1}^{N_{f}^{'}}\dfrac{(\sigma_{j}-m_{a}^{'})}{(-\sigma_{j}-m_{a}^{'})}=1
\end{aligned}
\end{equation}
We can rewrite the vacuum equation as two type vacuum equation equivalently after taking the square root of (\ref{21}) 
\begin{equation}\label{16}
\dfrac{(\sigma_{j}-\frac{m_{\text{adj}}}{2})}{(\sigma_{j}+\frac{m_{\text{adj}}}{2})}\prod_{j\neq k}^{N}\dfrac{(\sigma_{j}\pm \sigma_{k}-m_{\text{adj}})}{(-\sigma_{j}\pm \sigma_{k}-m_{\text{adj}})}\prod_{a=1}^{N_{f}}\dfrac{(\sigma_{j}-m_{a})}{(-\sigma_{j}-m_{a})}=1
\end{equation}
and
\begin{equation}\label{}
\dfrac{(\sigma_{j}-\frac{m_{\text{adj}}}{2})}{(\sigma_{j}+\frac{m_{\text{adj}}}{2})}\prod_{j\neq k}^{N}\dfrac{(\sigma_{j}\pm \sigma_{k}-m_{\text{adj}})}{(-\sigma_{j}\pm \sigma_{k}-m_{\text{adj}})}\prod_{a=1}^{N_{f}}\dfrac{(\sigma_{j}-m_{a})}{(-\sigma_{j}-m_{a})}=-1
\end{equation}
With the same dictionary (\ref{15}), we can choose one of the following boundary conditions
\begin{itemize}
\item \qquad $\xi_{+}=\dfrac{\eta}{2},\qquad \xi_{-}=0$
\item \qquad $\xi_{-}=\dfrac{\eta}{2},\qquad \xi_{+}=0$
\end{itemize}
to obtain Bethe/Gauge correspondence. We can see  the boundary condition parameters are different at the end of the spin chain in our correspondence. In \cite{KZ21}, the boundary condition of the Bethe ansatz equation (\ref{13}) is chosen as $\xi_{+}=\xi_{-}=0$.

For $\text{B}$-type theory, the vacuum equation is 
\begin{equation}\label{22}
\begin{aligned}
&\dfrac{(\sigma_{j}-m_{\text{adj}})^{4}}{(\sigma_{j}+m_{\text{adj}})^{4}}\prod_{j\neq k}^{N}\dfrac{(\sigma_{j}\pm \sigma_{k}-m_{\text{adj}})}{(-\sigma_{j}\pm \sigma_{k}-m_{\text{adj}})}\prod_{a=1}^{N_{f}}\dfrac{(\sigma_{j}-m_{a})}{(-\sigma_{j}-m_{a})}\\
&\times\prod_{j\neq k}^{N}\dfrac{(\sigma_{j}\pm \sigma_{k}-m_{\text{adj}})}{(-\sigma_{j}\pm \sigma_{k}-m_{\text{adj}})}\prod_{a=1}^{N_{f}^{'}}\dfrac{(\sigma_{j}-m_{a}^{'})}{(-\sigma_{j}-m_{a}^{'})}=1
\end{aligned}
\end{equation}
We can rewrite the vacuum equation as two type vacuum equation equivalently after taking the square root of (\ref{22})
\begin{equation}\label{40}
\dfrac{(\sigma_{j}-m_{\text{adj}})^{2}}{(\sigma_{j}+m_{\text{adj}})^{2}}\prod_{j\neq k}^{N}\dfrac{(\sigma_{j}\pm \sigma_{k}-m_{\text{adj}})}{(-\sigma_{j}\pm \sigma_{k}-m_{\text{adj}})}\prod_{a=1}^{N_{f}}\dfrac{(\sigma_{j}-m_{a})}{(-\sigma_{j}-m_{a})}=1
\end{equation}
and
\begin{equation}
\dfrac{(\sigma_{j}-m_{\text{adj}})^{2}}{(\sigma_{j}+m_{\text{adj}})^{2}}\prod_{j\neq k}^{N}\dfrac{(\sigma_{j}\pm \sigma_{k}-m_{\text{adj}})}{(-\sigma_{j}\pm \sigma_{k}-m_{\text{adj}})}\prod_{a=1}^{N_{f}^{'}}\dfrac{(\sigma_{j}-m_{a}^{'})}{(-\sigma_{j}-m_{a}^{'})}=1
\end{equation}
The same dictionary (\ref{15}) maps the vacuum equation (\ref{40}) to the Bethe ansatz equation (\ref{13}) with the boundary condition
\begin{equation}
 \xi_{+}=\xi_{-}=-\dfrac{\eta}{2}
\end{equation}
which is different from the boundary condition parameters for $\text{SO}(2N+1)$ supersymmetry gauge theory in \cite{KZ21}. The boundary parameters are the same at the two ends of the spin chain in our results. Just like the case in 3d, we choose $s_{1}=s_{2}=\dfrac{1}{2}$ and $\vartheta_{1}=\vartheta_{2}=0$, then we get the corresponding $m_{1}\longleftrightarrow - \eta$, $m_{2}\longleftrightarrow 0$ and
\begin{equation}
\dfrac{(\sigma_{j}-m_{1})}{(-\sigma_{j}-m_{1})}=\dfrac{(\sigma_{j}+\eta)}{(-\sigma_{j}+\eta)},\quad \dfrac{(\sigma_{j}-m_{2})}{(-\sigma_{j}-m_{2})}=-1
\end{equation}
The factor $\frac{(\sigma_{j}-m_{\text{adj}})^{2}}{(\sigma_{j}+m_{\text{adj}})^{2}}$ is simplified to $\frac{(\sigma_{j}-m_{\text{adj}})}{(\sigma_{j}+m_{\text{adj}})}$. Then the vacuum equation (\ref{40}) is the same as the vacuum equation of 2d $\text{B}$-type gauge theory in \cite{KZ21}. And the dictionary is changed to 
\begin{equation}\label{}
\begin{aligned}
&u \longleftrightarrow \sigma ,\quad \eta \longleftrightarrow m_{\text{adj}}\\
&M \longleftrightarrow N,\quad 2L\longleftrightarrow N_{f}-2\\
&\{-\eta s_{a}-\dfrac{\eta}{2} +\vartheta_{a},-\eta s_{a}+\dfrac{ \eta}{2} -\vartheta_{a}\} \longleftrightarrow  m_{a}
 \end{aligned}
\end{equation}
So the vacuum equation of 2d $\text{B}$-type supersymmetry gauge theory in \cite{KZ21} is a special case of ours.

For $\text{BC}_{N}$-type gauge groups, the vacuum equations degenerate to 
\begin{equation}\label{41}
\begin{aligned}
&\dfrac{(\sigma_{j}-m_{\text{adj}})^{4}}{(\sigma_{j}+m_{\text{adj}})^{4}}\prod_{j\neq k}^{N}\dfrac{(\sigma_{j}\pm \sigma_{k}-m_{\text{adj}})}{(-\sigma_{j}\pm \sigma_{k}-m_{\text{adj}})}\prod_{a=1}^{N_{f}}\dfrac{(\sigma_{j}-m_{a})}{\text{sin}(-\sigma_{j}-m_{a})}\\
&\times \dfrac{(2\sigma_{j}-m_{\text{adj}})^{2}}{(2\sigma_{j}+m_{\text{adj}})^{2}}\prod_{j\neq k}^{N}\dfrac{(\sigma_{j}\pm \sigma_{k}-m_{\text{adj}})}{(-\sigma_{j}\pm \sigma_{k}-m_{adj})}\prod_{a=1}^{N_{f}^{'}}\dfrac{(\sigma_{j}-m_{a}^{'})}{(-\sigma_{j}-m_{a}^{'})}=1
\end{aligned}
\end{equation}
Equivalently, the vacuum equation can be written as two type vacuum equation after taking the square root of (\ref{41})
\begin{equation}\label{42}
\dfrac{(\sigma_{j}-\frac{m_{\text{adj}}}{2})}{(\sigma_{j}+\frac{m_{\text{adj}}}{2})}\dfrac{(\sigma_{j}-m_{\text{adj}})^{2}}{(\sigma_{j}+m_{\text{adj}})^{2}}\prod_{j\neq k}^{N}\dfrac{(\sigma_{j}\pm \sigma_{k}-m_{\text{adj}})}{(-\sigma_{j}\pm \sigma_{k}-m_{\text{adj}})}\prod_{a=1}^{N_{f}}\dfrac{(\sigma_{j}-m_{a})}{(-\sigma_{j}-m_{a})}=1
\end{equation}
and
\begin{equation}\label{}
\dfrac{(\sigma_{j}-\frac{m_{\text{adj}}}{2})}{(\sigma_{j}+\frac{m_{\text{adj}}}{2})}\dfrac{(\sigma_{j}-m_{\text{adj}})^{2}}{(\sigma_{j}+m_{\text{adj}})^{2}}\prod_{j\neq k}^{N}\dfrac{(\sigma_{j}\pm \sigma_{k}-m_{\text{adj}})}{(-\sigma_{j}\pm \sigma_{k}-m_{\text{adj}})}\prod_{a=1}^{N_{f}}\dfrac{(\sigma_{j}-m_{a})}{(-\sigma_{j}-m_{a})}=-1
\end{equation}
If we choose $s_{1}=s_{2}=s_{3}=s_{4}=\dfrac{1}{2}$ and $\vartheta_{1}=\vartheta_{2}=\vartheta_{3}=\vartheta_{4}=0$, then we get the corresponding $m_{1}=m_{2}\longleftrightarrow -\eta$, $m_{2}=m_{4} \longleftrightarrow 0$ and
 \begin{equation}
 \dfrac{(\sigma_{j}-m_{1})}{(-\sigma_{j}-m_{1})}=\dfrac{(\sigma_{j}+\pi \eta)}{(-\sigma_{j}+\pi\eta)},\quad  \dfrac{(\sigma_{j}-m_{2})}{(-\sigma_{j}-m_{2})}=-1
\end{equation}
\begin{equation}
 \dfrac{(\sigma_{j}-m_{3})}{(-\sigma_{j}-m_{3})}=\dfrac{(\sigma_{j}+\pi \eta)}{(-\sigma_{j}+\pi\eta)},\quad  \dfrac{(\sigma_{j}-m_{4})}{(-\sigma_{j}-m_{4})}=-1
\end{equation}
It is easy to prove that the factor $\frac{(\sigma_{j}-\frac{m_{\text{adj}}}{2})}{(\sigma_{j}+\frac{m_{\text{adj}}}{2})}\frac{(\sigma_{j}-m_{\text{adj}})^{2}}{(\sigma_{j}+m_{\text{adj}})^{2}}$ is reduced to $\frac{(\sigma_{j}-\frac{m_{\text{adj}}}{2})}{(\sigma_{j}+\frac{m_{\text{adj}}}{2})}$,  and the vacuum equation (\ref{42}) can be mapped to the Bethe ansatz equation (\ref{13}) with the boundary condition
\begin{itemize}
\item \qquad $\xi_{+}=\dfrac{\eta}{2},\qquad \xi_{-}=0$
\item \qquad $\xi_{-}=\dfrac{\eta}{2},\qquad \xi_{+}=0$
\end{itemize}
And the dictionary is changed into 
\begin{equation}\label{}
\begin{aligned}
&u \longleftrightarrow \sigma ,\quad \eta \longleftrightarrow m_{\text{adj}}\\
&M \longleftrightarrow N,\quad 2L\longleftrightarrow N_{f}-4\\
&\{-\eta s_{a}-\dfrac{\eta}{2} +\vartheta_{a},-\eta s_{a}+\dfrac{ \eta}{2} -\vartheta_{a}\} \longleftrightarrow  m_{a}
 \end{aligned}
\end{equation} 
Similar to 3d $\text{BC}_{N}$ theory, we let some parameters take fixed values to get Bethe/Gauge correspondence. 

\section{Exceptional Lie algebras}\label{d}

Except for classical Lie algebra $\text{A}_{N}$, $\text{B}_{N}$, $\text{C}_{N}$, $\text{D}_{N}$, we also consider exceptional Lie algebras $\text{E}_{6},\text{E}_{7},\text{E}_{8},\text{F}_{4}$ and $\text{G}_{2}$. The process for calculating $W_{\text{eff}}^{3d}(\sigma,m)$ is the same to the classical Lie algebra. The effective potential is given by (\ref{48}).

\subsection{$\text{F}_{4}$-type}

The root system of $\text{F}_{4}$ consists of all $\pm \sigma_{i}$, all $\pm(\sigma_{i}\pm \sigma_{j})$, $i\neq j$, as well as all $\pm \frac{1}{2}(\sigma_{1}\pm \sigma_{2}\pm \sigma_{3}\pm \sigma_{4})$, where the signs may be chosen independently \cite{Hum72}. The effective superpotential is given by
\begin{equation}\label{}
\begin{aligned}
W_{\text{eff}}^{3d}(\sigma,m)=&-\dfrac{2}{\beta_{2}}\sum_{j\neq k}^{4}\text{Li}_{2}(e^{i(\pm\sigma_{j}\pm\sigma_{k})})+\dfrac{1}{2\beta_{2}}\sum_{j\neq k}^{4}(\pm\sigma_{j}\pm \sigma_{k})^{2}\\
&+\dfrac{2}{\beta_{2}}\sum_{j\neq k}^{4}\text{Li}_{2}(e^{-i(\pm\sigma_{j}\pm\sigma_{k})-im_{\text{adj}}})-\dfrac{1}{2\beta_{2}}\sum_{j\neq k}^{4}(\pm\sigma_{j}\pm \sigma_{k}+m_{\text{adj}})^{2}\\
&-\dfrac{4}{\beta_{2}}\sum_{j=1}^{4}\text{Li}_{2}(e^{\pm i \sigma_{j}})+\dfrac{1}{\beta_{2}}\sum_{j=1}^{4}(\sigma_{j})^{2}\\
&+\dfrac{4}{\beta_{2}}\sum_{j=1}^{4}\text{Li}_{2}(e^{\pm i\sigma_{j}-im_{\text{adj}}})-\dfrac{1}{\beta_{2}}\sum_{j=1}^{4}(\sigma_{j}\pm m_{\text{adj}})^{2}\\
&-\dfrac{4}{\beta_{2}}\text{Li}_{2}(e^{i(\pm \frac{1}{2}(\sigma_{1}\pm \sigma_{2}\pm \sigma_{3}\pm \sigma_{4}))})+\dfrac{1}{4\beta_{2}}(\sigma_{1}\pm \sigma_{2}\pm \sigma_{3}\pm\sigma_{4})^{2}\\
&+\dfrac{4}{\beta_{2}}\text{Li}_{2}(e^{-i(\pm \frac{1}{2}(\sigma_{1}\pm \sigma_{2}\pm \sigma_{3}\pm \sigma_{4}))-im_{\text{adj}}})-\dfrac{1}{4\beta_{2}}(\sigma_{1}\pm \sigma_{2}\pm \sigma_{3}\pm\sigma_{4}\pm m_{\text{adj}})^{2}\\
&+\dfrac{1}{\beta_{2}}\sum_{j=1}^{4}\sum_{a=1}^{N_{f}}\text{Li}_{2}(e^{-i(\pm \sigma_{j}+m_{a})})-\dfrac{1}{4\beta_{2}}\sum_{j=1}^{4}\sum_{a=1}^{N_{f}}(\pm \sigma_{j}+m_{a})^{2}\\
&+\dfrac{1}{\beta_{2}}\sum_{j=1}^{4}\sum_{a=1}^{N_{f}^{'}}\text{Li}_{2}(e^{i(\pm \sigma_{j}-m_{a}^{'})})-\dfrac{1}{4\beta_{2}}\sum_{j=1}^{4}\sum_{a=1}^{N_{f}^{'}}(\pm \sigma_{j}-m_{a}^{'})^{2}
\end{aligned}
\end{equation}
By using the equation (\ref{100}), we write the corresponding vacuum equations as
\begin{equation}\label{207}
\begin{aligned}
&\dfrac{\text{sin}^{4}(\sigma_{j}-m_{\text{adj}})}{\text{sin}^{4}(\sigma_{j}+m_{\text{adj}})}\prod_{k\neq j}^{4}\dfrac{\text{sin}^{2}(\sigma_{j}\pm \sigma_{k}-m_{\text{adj}})}{\text{sin}^{2}(-\sigma_{j}\pm \sigma_{k}-m_{\text{adj}})}\dfrac{\text{sin}^{2}(\frac{1}{2}\sigma_{j}\pm \frac{1}{2}\sum_{k\neq j}^{4}\sigma_{k}-m_{\text{adj}})}{\text{sin}^{2}(-\frac{1}{2}\sigma_{j}\pm \frac{1}{2}\sum_{k\neq j}^{4}\sigma_{k}-m_{\text{adj}})}\\
&\times \prod_{a=1}^{N_{f}}\dfrac{\text{sin}(\sigma_{j}-m_{a})}{\text{sin}(-\sigma_{j}-m_{a})}\prod_{a=1}^{N_{f}^{'}}\dfrac{\text{sin}(\sigma_{j}-m_{a}^{'})}{\text{sin}(-\sigma_{j}-m_{a}^{'})}=1
\end{aligned}
\end{equation}
where $j,k=1,2,3,4$. There are 4 vacuum equations for $\text{F}_{4}$ gauge theory totally. Considering $N_{f}=N_{f}^{'}$, we rewrite the vacuum equation with the square root of (\ref{207})
\begin{equation}
\begin{aligned}
&\dfrac{\text{sin}^{2}(\sigma_{j}-m_{\text{adj}})}{\text{sin}^{2}(\sigma_{j}+m_{\text{adj}})}\prod_{k\neq j}^{4}\dfrac{\text{sin}(\sigma_{j}\pm \sigma_{k}-m_{\text{adj}})}{\text{sin}(-\sigma_{j}\pm \sigma_{k}-m_{\text{adj}})}\dfrac{\text{sin}(\frac{1}{2}\sigma_{j}\pm \frac{1}{2}\sum_{k\neq j}^{4}\sigma_{k}-m_{\text{adj}})}{\text{sin}(-\frac{1}{2}\sigma_{j}\pm\frac{1}{2} \sum_{k\neq j}^{4}\sigma_{k}-m_{\text{adj}})}\\
&\times \prod_{a=1}^{N_{f}}\dfrac{\text{sin}(\sigma_{j}-m_{a})}{\text{sin}(-\sigma_{j}-m_{a})}=1
\end{aligned}
\end{equation}
and
\begin{equation}
\begin{aligned}
&\dfrac{\text{sin}^{2}(\sigma_{j}-m_{\text{adj}})}{\text{sin}^{2}(\sigma_{j}+m_{\text{adj}})}\prod_{k\neq j}^{4}\dfrac{\text{sin}(\sigma_{j}\pm \sigma_{k}-m_{\text{adj}})}{\text{sin}(-\sigma_{j}\pm \sigma_{k}-m_{\text{adj}})}\dfrac{\text{sin}(\frac{1}{2}\sigma_{j}\pm \frac{1}{2}\sum_{k\neq j}^{4}\sigma_{k}-m_{\text{adj}})}{\text{sin}(-\frac{1}{2}\sigma_{j}\pm \frac{1}{2}\sum_{k\neq j}^{4}\sigma_{k}-m_{\text{adj}})}\\
&\times \prod_{a=1}^{N_{f}}\dfrac{\text{sin}(\sigma_{j}-m_{a})}{\text{sin}(-\sigma_{j}-m_{a})}=-1
\end{aligned}
\end{equation}
Inspired by the results with classical Lie algebras, we may sure that, one must match the two types of vacuum equations with certain integrable system, maybe a spin chain with given representation. However, to our knowledge, there are no explicit Bethe equations in existing articles corresponding to these equations.

\subsection{$\text{G}_{2}$-type}

Let Euclidean space $\mathbf{E}$ be the $2$-dimensional subspace of $R^{3}$ orthogonal to $\sigma_{1}+\sigma_{2}+\sigma_{3}$. It is a constraint condition on roots. Then the root system of $\text{G}_{2}$ consists of \cite{Hum72}
$$\pm\left\{\sigma_{1}-\sigma_{2},\sigma_{2}-\sigma_{3},\sigma_{1}-\sigma_{3},2\sigma_{1}-\sigma_{2}-\sigma_{3},2\sigma_{2}-\sigma_{1}-\sigma_{3},2\sigma_{3}-\sigma_{1}-\sigma_{2}\right\}$$
The effective superpotential is given by
\begin{equation}\label{}
\begin{aligned}
W_{\text{eff}}^{3d}(\sigma,m)=&-\dfrac{2}{\beta_{2}}\sum_{j<k}\text{Li}_{2}(e^{\pm i(\sigma_{j}-\sigma_{k})})+\dfrac{2}{\beta_{2}}\sum_{j< k}(\sigma_{j}-\sigma_{k})^{2}\\
&+\dfrac{2}{\beta_{2}}\sum_{j< k}\text{Li}_{2}(e^{-\pm i(\sigma_{j}-\sigma_{k})-im_{\text{adj}}})-\dfrac{1}{2\beta_{2}}\sum_{j<q k}(\pm\sigma_{j}- \sigma_{k}+m_{\text{adj}})^{2}\\
&-\dfrac{2}{3\beta_{2}}\text{Li}_{2}(e^{\pm i (2\sigma_{1}-\sigma_{2}-\sigma_{3})})+\dfrac{1}{9\beta_{2}}(2\sigma_{1}-\sigma_{2}-\sigma_{3})^{2}\\
&+\dfrac{2}{3\beta_{2}}\text{Li}_{2}(e^{\pm i(2\sigma_{1}-\sigma_{2}-\sigma_{3})-im_{\text{adj}}})-\dfrac{1}{6\beta_{2}}(2\sigma_{1}-\sigma_{2}-\sigma_{3}\pm m_{\text{adj}})^{2}\\
&-\dfrac{2}{3\beta_{2}}\text{Li}_{2}(e^{\pm i (2\sigma_{2}-\sigma_{1}-\sigma_{3})})+\dfrac{1}{6\beta_{2}}(2\sigma_{2}-\sigma_{1}-\sigma_{3})^{2}\\
&+\dfrac{2}{3\beta_{2}}\text{Li}_{2}(e^{\pm i(2\sigma_{2}-\sigma_{1}-\sigma_{3})-im_{\text{adj}}})-\dfrac{1}{9\beta_{2}}(2\sigma_{2}-\sigma_{1}-\sigma_{3}\pm m_{\text{adj}})^{2}\\
&-\dfrac{2}{3\beta_{2}}\text{Li}_{2}(e^{\pm i (2\sigma_{3}-\sigma_{1}-\sigma_{2})})+\dfrac{1}{6\beta_{2}}(2\sigma_{3}-\sigma_{1}-\sigma_{2})^{2}\\
&+\dfrac{2}{3\beta_{2}}\text{Li}_{2}(e^{\pm i(2\sigma_{3}-\sigma_{1}-\sigma_{2})-im_{\text{adj}}})-\dfrac{1}{9\beta_{2}}(2\sigma_{3}-\sigma_{1}-\sigma_{2}\pm m_{\text{adj}})^{2}\\
&+\dfrac{1}{\beta_{2}}\sum_{j=1}^{3}\sum_{a=1}^{N_{f}}\text{Li}_{2}(e^{-i(\pm \sigma_{j}+m_{a})})-\dfrac{1}{4\beta_{2}}\sum_{j=1}^{4}\sum_{a=1}^{N_{f}}(\pm \sigma_{j}+m_{a})^{2}\\
&+\dfrac{1}{\beta_{2}}\sum_{j=1}^{3}\sum_{a=1}^{N_{f}^{'}}\text{Li}_{2}(e^{i(\pm \sigma_{j}-m_{a}^{'})})-\dfrac{1}{4\beta_{2}}\sum_{j=1}^{4}\sum_{a=1}^{N_{f}}(\pm \sigma_{j}-m_{a}^{'})^{2}
\end{aligned}
\end{equation}
By using the formula (\ref{100}), we can get the following three vacuum equations. For $\sigma_{1}$, the vacuum equation can be written as
\begin{equation}\label{208}
\begin{aligned}
&\dfrac{\text{sin}^{2}(\sigma_{1}-\sigma_{2}-m_{\text{adj}})}{\text{sin}^{2}(-\sigma_{1}-\sigma_{2}-m_{\text{adj}})}\dfrac{\text{sin}^{2}(\sigma_{1}-\sigma_{3}-m_{\text{adj}})}{\text{sin}^{2}(-\sigma_{1}-\sigma_{3}-m_{\text{adj}})}\dfrac{\text{sin}^{\frac{4}{3}}(2\sigma_{1}-\sigma_{2}-\sigma_{3}-m_{\text{adj}})}{\text{sin}^{\frac{4}{3}}(-2\sigma_{1}-\sigma_{2}-\sigma_{3}-m_{\text{adj}})}\\
&\times \dfrac{\text{sin}^{\frac{2}{3}}(\sigma_{1}-2\sigma_{2}+\sigma_{3}-m_{\text{adj}})}{\text{sin}^{\frac{2}{3}}(-\sigma_{1}-2\sigma_{2}+\sigma_{3}-m_{\text{adj}})}\dfrac{\text{sin}^{\frac{2}{3}}(\sigma_{1}-2\sigma_{3}+\sigma_{2}-m_{\text{adj}})}{\text{sin}^{\frac{2}{3}}(-\sigma_{1}-2\sigma_{3}+\sigma_{2}-m_{\text{adj}})}\\
&\times \prod_{a=1}^{N_{f}}\dfrac{\text{sin}(\sigma_{1}-m_{a})}{\text{sin}(-\sigma_{1}-m_{a})}\prod_{a=1}^{N_{f}^{'}}\dfrac{\text{sin}(\sigma_{1}-m_{a}^{'})}{\text{sin}(-\sigma_{1}-m_{a}^{'})}=1
\end{aligned}
\end{equation}
For $\sigma_{2}$, the vacuum equation is given by
\begin{equation}
\begin{aligned}
&\dfrac{\text{sin}^{2}(\sigma_{2}-\sigma_{1}-m_{\text{adj}})}{\text{sin}^{2}(-\sigma_{2}-\sigma_{1}-m_{\text{adj}})}\dfrac{\text{sin}^{2}(\sigma_{2}-\sigma_{3}-m_{\text{adj}})}{\text{sin}^{2}(-\sigma_{2}-\sigma_{3}-m_{\text{adj}})}\dfrac{\text{sin}^{\frac{4}{3}}(2\sigma_{2}-\sigma_{1}-\sigma_{3}-m_{\text{adj}})}{\text{sin}^{\frac{4}{3}}(-2\sigma_{2}-\sigma_{1}-\sigma_{3}-m_{\text{adj}})}\\
&\times \dfrac{\text{sin}^{\frac{2}{3}}(\sigma_{2}-2\sigma_{1}+\sigma_{3}-m_{\text{adj}})}{\text{sin}^{\frac{2}{3}}(-\sigma_{2}-2\sigma_{1}+\sigma_{3}-m_{\text{adj}})}\dfrac{\text{sin}^{\frac{2}{3}}(\sigma_{2}-2\sigma_{3}+\sigma_{1}-m_{\text{adj}})}{\text{sin}^{\frac{2}{3}}(-\sigma_{2}-2\sigma_{3}+\sigma_{1}-m_{\text{adj}})}\\
&\times \prod_{a=1}^{N_{f}}\dfrac{\text{sin}(\sigma_{2}-m_{a})}{\text{sin}(-\sigma_{2}-m_{a})}\prod_{a=1}^{N_{f}^{'}}\dfrac{\text{sin}(\sigma_{2}-m_{a}^{'})}{\text{sin}(-\sigma_{2}-m_{a}^{'})}=1
\end{aligned}
\end{equation}
For $\sigma_{3}$, the vacuum equation reads 
\begin{equation}
\begin{aligned}
&\dfrac{\text{sin}^{2}(\sigma_{3}-\sigma_{1}-m_{\text{adj}})}{\text{sin}^{2}(-\sigma_{3}-\sigma_{1}-m_{\text{adj}})}\dfrac{\text{sin}^{2}(\sigma_{3}-\sigma_{2}-m_{\text{adj}})}{\text{sin}^{2}(-\sigma_{3}-\sigma_{2}-m_{\text{adj}})}\dfrac{\text{sin}^{\frac{4}{3}}(2\sigma_{3}-\sigma_{1}-\sigma_{2}-m_{\text{adj}})}{\text{sin}^{\frac{4}{3}}(-2\sigma_{3}-\sigma_{1}-\sigma_{2}-m_{\text{adj}})}\\
&\times \dfrac{\text{sin}^{\frac{2}{3}}(\sigma_{3}-2\sigma_{1}+\sigma_{2}-m_{\text{adj}})}{\text{sin}^{\frac{2}{3}}(-\sigma_{3}-2\sigma_{1}+\sigma_{2}-m_{\text{adj}})}\dfrac{\text{sin}^{\frac{2}{3}}(\sigma_{3}-2\sigma_{2}+\sigma_{1}-m_{\text{adj}})}{\text{sin}^{\frac{2}{3}}(-\sigma_{3}-2\sigma_{2}+\sigma_{1}-m_{\text{adj}})}\\
&\times \prod_{a=1}^{N_{f}}\dfrac{\text{sin}(\sigma_{3}-m_{a})}{\text{sin}(-\sigma_{3}-m_{a})}\prod_{a=1}^{N_{f}^{'}}\dfrac{\text{sin}(\sigma_{3}-m_{a}^{'})}{\text{sin}(-\sigma_{3}-m_{a}^{'})}=1
\end{aligned}
\end{equation}
For convenience, we only write the square rooted vacuum equation of (\ref{208})
\begin{equation}
\begin{aligned}
&\dfrac{\text{sin}(\sigma_{1}-\sigma_{2}-m_{\text{adj}})}{\text{sin}(-\sigma_{1}-\sigma_{2}-m_{\text{adj}})}\dfrac{\text{sin}(\sigma_{1}-\sigma_{3}-m_{\text{adj}})}{\text{sin}(-\sigma_{1}-\sigma_{3}-m_{\text{adj}})}\dfrac{\text{sin}^{\frac{2}{3}}(2\sigma_{1}-\sigma_{2}-\sigma_{3}-m_{\text{adj}})}{\text{sin}^{\frac{2}{3}}(-2\sigma_{1}-\sigma_{2}-\sigma_{3}-m_{\text{adj}})}\\
&\times \dfrac{\text{sin}^{\frac{1}{3}}(\sigma_{1}-2\sigma_{2}+\sigma_{3}-m_{\text{adj}})}{\text{sin}^{\frac{1}{3}}(-\sigma_{1}-2\sigma_{2}+\sigma_{3}-m_{\text{adj}})}\dfrac{\text{sin}^{\frac{1}{3}}(\sigma_{1}-2\sigma_{3}+\sigma_{2}-m_{\text{adj}})}{\text{sin}^{\frac{1}{3}}(-\sigma_{1}-2\sigma_{3}+\sigma_{2}-m_{\text{adj}})}\\
&\times \prod_{a=1}^{N_{f}}\dfrac{\text{sin}(\sigma_{1}-m_{a})}{\text{sin}(-\sigma_{1}-m_{a})}=1
\end{aligned}
\end{equation}
and
\begin{equation}
\begin{aligned}
&\dfrac{\text{sin}(\sigma_{1}-\sigma_{2}-m_{\text{adj}})}{\text{sin}(-\sigma_{1}-\sigma_{2}-m_{\text{adj}})}\dfrac{\text{sin}(\sigma_{1}-\sigma_{3}-m_{\text{adj}})}{\text{sin}(-\sigma_{1}-\sigma_{3}-m_{\text{adj}})}\dfrac{\text{sin}^{\frac{2}{3}}(2\sigma_{1}-\sigma_{2}-\sigma_{3}-m_{\text{adj}})}{\text{sin}^{\frac{2}{3}}(-2\sigma_{1}-\sigma_{2}-\sigma_{3}-m_{\text{adj}})}\\
&\times \dfrac{\text{sin}^{\frac{1}{3}}(\sigma_{1}-2\sigma_{2}+\sigma_{3}-m_{\text{adj}})}{\text{sin}^{\frac{1}{3}}(-\sigma_{1}-2\sigma_{2}+\sigma_{3}-m_{\text{adj}})}\dfrac{\text{sin}^{\frac{1}{3}}(\sigma_{1}-2\sigma_{3}+\sigma_{2}-m_{\text{adj}})}{\text{sin}^{\frac{1}{3}}(-\sigma_{1}-2\sigma_{3}+\sigma_{2}-m_{\text{adj}})}\\
&\times \prod_{a=1}^{N_{f}}\dfrac{\text{sin}(\sigma_{1}-m_{a})}{\text{sin}(-\sigma_{1}-m_{a})}=-1
\end{aligned}
\end{equation}
Also, the two types of vacuum equations can be corresponded to the Bethe equations of certain valued spin chain.

\subsection{$\text{E}_{8}$-type}

We know that $E_{6}$, $E_{7}$ can be identified canonically with subsystems of $E_{8}$, so it is suffices to construct $E_{8}$. The root system of $E_{8}$ consists of the obvious vectors $\pm(\sigma_{i}\pm \sigma_{j})$, $i\neq j$, $i,j=1,\cdots,8$, along with the less obvious ones $\frac{1}{2}\sum_{i=1}^{8}(-1)^{\kappa(i)}\sigma_{i}$ (where the $\kappa(i)=0,1$, add up to an even integer) \cite{Hum72}. The effective superpotential is given by

\begin{equation}\label{}
\begin{aligned}
W_{\text{eff}}^{3d}(\sigma,m)=&-\dfrac{2}{\beta_{2}}\sum_{j\neq k}^{8}\text{Li}_{2}(e^{i(\pm\sigma_{j}\pm\sigma_{k})})+\dfrac{1}{2\beta_{2}}\sum_{j\neq k}^{8}(\pm\sigma_{j}\pm \sigma_{k})^{2}\\
&+\dfrac{2}{\beta_{2}}\sum_{j\neq k}^{8}\text{Li}_{2}(e^{-i(\pm\sigma_{j}\pm\sigma_{k})-im_{\text{adj}}})-\dfrac{1}{2\beta_{2}}\sum_{j\neq k}^{8}(\pm\sigma_{j}\pm \sigma_{k}+m_{\text{adj}})^{2}\\
&-\dfrac{2}{\beta_{2}}\text{Li}_{2}(e^{i(\frac{1}{2}\sum_{j=1}^{8}(-1)^{\kappa(j)}\sigma_{j})})+\dfrac{1}{2\beta_{2}}(\sum_{j=1}^{8}(-1)^{\kappa(j)}\sigma_{j})^{2}\\
&+\dfrac{2}{\beta_{2}}\text{Li}_{2}(e^{-i(\frac{1}{2}\sum_{j=1}^{8}(-1)^{\kappa(j)}\sigma_{j})-im_{\text{adj}}})-\dfrac{1}{2\beta_{2}}(\sum_{j=1}^{8}(-1)^{\kappa(j)}\sigma_{j}+m_{\text{adj}})^{2}\\
&+\dfrac{1}{\beta_{2}}\sum_{j=1}^{8}\sum_{a=1}^{N_{f}}\text{Li}_{2}(e^{-i(\pm \sigma_{j}+m_{a})})-\dfrac{1}{4\beta_{2}}\sum_{j=1}^{8}\sum_{a=1}^{N_{f}}(\pm \sigma_{j}+m_{a})^{2}\\
&+\dfrac{1}{\beta_{2}}\sum_{j=1}^{8}\sum_{a=1}^{N_{f}^{'}}\text{Li}_{2}(e^{i(\pm \sigma_{j}-m_{a}^{'})})-\dfrac{1}{4\beta_{2}}\sum_{j=1}^{8}\sum_{a=1}^{N_{f}^{'}}(\pm \sigma_{j}-m_{a}^{'})^{2}
\end{aligned}
\end{equation}
Different $\sum_{j=1}^{8}(-1)^{\kappa(j)}\sigma_{j}$ corresponds to different vacuum equation. We are not going to write it all out in order to save space. The selection of root systems is not unique. We choose only one type here. One can use (\ref{100}) to get the explicit vacuum equations. The root system $\text{E}_{8}$ has 240 roots , the root system $\text{E}_{7}$ has 126 roots and the root system $\text{E}_{6}$ has 72 roots. 

Because of the difference between the root system of classical Lie algebra and the root system of exceptional Lie algebra, we can see that the vacuum equation of exceptional Lie algebra cannot correspond to the Bethe ansatz equation of $\text{XXZ}$ spin chain with diagonal boundary condition with easy performances. We only give the effective superpotential and vacuum equation of exceptional Lie algebra here. For the corresponding integrable system, as far as we know, there are no results in the current literature yet.

\section{Conclusions and future prospects}\label{e}

In this article, we find more correspondence between supersymmetry gauge theory and integrable system with respect to the results in \cite{KZ21}, as we changed 
the representation and consequently the expression of the superpotential. We generalized the Bethe/Gauge correspondences first proposed in \cite{NS09b,NS09c} for $A$-type gauge theories to $\text{BCD}$-type gauge groups. We saw that the correspondence worked perfectly not only for 2d gauge theories but also for 3d. Furthermore, we found that we can always choose diagonal boundary conditions for open $\text{XXZ}$ spin chain with boundary parameters $\xi=\pm\frac{\eta}{2},0, \infty$ or $\frac{2k+1}{2}$ to realize the vacuum equation of gauge theory from the Bethe ansatz equation. In addition to the classical Lie algebras, we also considered a new $\text{BC}_{N}$-type gauge theory. In the case of certain parameters with specific values, the vacuum equations of $\text{BC}_{N}$-type gauge theory can also be identified as the Bethe ansatz equations of open $\text{XXZ}$ spin chain with diagonal boundary condition. We also gave the effective superpotential of the exceptional Lie algebra $\text{E}_{6,7,8}$, $\text{F}_{4}$ and $\text{G}_{2}$ with our new representation. Specially, we give the vacuum equations of $\text{F}_{4}$ and $\text{G}_{2}$. However, the corresponding integrable system has not been found yet.

Another possible way to solve the problems of $\text{C}$-type gauge theory is to change the boundary conditions of the open spin chain with new Bethe equation. Then it can be checked out that, whether such Bethe equation works on the vacuum equation of $\text{C}$-type gauge theory such as in \cite{KZ21}. The Bethe ansatz equation of $\text{SU}(3)$ spin chain with generic boundary condition has been calculated in \cite{Sun18}. With two gauge nodes being different classical Lie groups, we also consider a higher level $\text{A}_{2}$ quiver gauge theory and higher rank $sl_{3}$ spin chain model in \cite{DZ23}. And we desire the correspondence also materialize with our new methods. We have known the exact partition function of 4d $\mathcal{R}=1$ theories on $D^{2}\times T^{2}$ \cite{LNP19}, and the Bethe ansatz equation of open $\text{XYZ}$ spin chains \cite{YZ06,FHSY96}. So we can further calculate the effective superpotential and the vacuum equation of 4d $\mathcal{N}=1$ $\text{BCD}$-type gauge theories on $D^{2}\times T^{2}$. And we expect that the Bethe/Gauge correspondence can be uplifted to 4d $\mathcal{N}=1$ gauge theories. If one considers the quiver gauge theories, one has to consider the 5d (or 6d) quiver gauge theories \cite{NPS18,Nek04,NP12}, and explored a particular formalism for the systematic try out of the universal part of the superpotential $\mathcal{W}(\mathfrak{a},\mathrm{m};\mathfrak{q},\epsilon)$ recently, without considering the choices of boundary conditions.

\acknowledgments
The financial support from the Natural Science Foundation of China (NSFC, Grants 11775299) is gratefully acknowledged from one of the authors (Ding).



\bibliographystyle{plain}

\end{document}